# RePrompt: Automatic Prompt Editing to Refine AI-Generative Art Towards Precise Expressions

Yunlong Wang*
Institute of High Performance Computing (IHPC), Agency for Science, Technology and Research (A*STAR), Singapore
wang_yunlong@ihpc.a-star.edu.sg

Shuyuan Shen
National University of Singapore, Singapore
e0950159@u.nus.edu

Brian Y. Lim
Department of Computer Science, National University of Singapore, Singapore
brianlim@comp.nus.edu.sg

## ABSTRACT

Generative AI models have shown impressive ability to produce images with text prompts, which could benefit creativity in visual art creation and self-expression. However, it is unclear how precisely the generated images express contexts and emotions from the input texts. We explored the emotional expressiveness of AI-generated images and developed RePrompt, an automatic method to refine text prompts toward precise expression of the generated images. Inspired by crowdsourced editing strategies, we curated intuitive text features, such as the number and concreteness of nouns, and trained a proxy model to analyze the feature effects on the AI-generated image. With model explanations of the proxy model, we curated a rubric to adjust text prompts to optimize image generation for precise emotion expression. We conducted simulation and user studies, which showed that RePrompt significantly improves the emotional expressiveness of AI-generated images, especially for negative emotions.

*The work was primarily done when the author was working at the National University of Singapore.

To appear in *Proceedings of the 2023 CHI Conference on Human Factors in Computing Systems (CHI '23), April 23–28, 2023, Hamburg, Germany*

https://doi.org/10.1145/3544548.3581402

## 1 INTRODUCTION

Artificial Intelligence (AI) has provided many benefits, from reducing labor work, supporting data analysis, content retrieving and recommendation, to even content generation in creative work. The advancement of pre-trained language-image cross-modal models (e.g., CLIP [60] and ALIGN [31]) has fueled generative models (e.g., GANs [8, 33] and Diffusion Modes [66]) to produce images using free-text prompts. These text-to-image generative AI models (e.g., [19, 51, 61, 67]) enable laypersons to quickly create visual artworks of any idea that can be described verbally. This technology might benefit people by supporting creativity and emotion expression. In the HCI community, researchers have been exploring how these black-box AI models work [41] and how they can be used to help users [43].

Prior works have primarily used a constrained set of keywords as prompts to study the behavior and prompt engineering of these AI models [41, 42]. For instance, Liu and Chilton's study [41], with prompts composed of 12 subjects (e.g., love, woman, and tree) and 12 styles (e.g., Surrealism, cyberpunk, and Disney), suggested that different rephrasing of a prompt using the same keywords did not yield significantly different image generations. Oppenlaender analyzed practitioners' posts in online communities around text-based generative art and identified five types of prompt modifiers (i.e., subject terms, style modifiers, quality boosters, repetitions, and magic terms) that users primarily used to improve image generations [57]. However, the feasibility of using text-to-image generative models for emotion expression remains unclear [22]. Inspired by these pioneering works, we further investigated prompt engineering using more complex prompts with the use case of emotion expression. We are especially interested in whether AI-generated images align with emotions conveyed in text prompts.

Our rationale for exploring emotional expression with AI is twofold: 1) emotional intelligence is arguably an indispensable aspect of judging the intelligence of AI [90], and 2) humans have an innate need to express their emotions. For example, visual art has been essential to communicate feelings, emotions, and experiences beyond the written word. In the daily scenarios of emotion expression and sharing, e.g., messaging to friends or posting on social media, we often directly describe our emotions with some context information. We are interested in bridging these scenarios with text-to-image generative AI to enable a more natural transmission from emotion expression using words to that using images. Therefore, the prompts in our study differ from the common descriptive prompts for the generative models. Using natural expressions as prompts connects different modalities (e.g., speech and chat) to text-to-image models. As shown in Figure 1a, one could express a sad feeling by writing "my best friend will be going to school in another country for 4 years" and a generative AI would generate a matching emotional image. However, the generated images might not capture the emotion well. A person could rewrite the emotional text but that may not be optimized for the generative AI (Figure 1b). Instead, we propose an automatic method to edit the text prompt so that the AI-generated images more precisely express the meaning and emotion in the original text (Figure 1c).

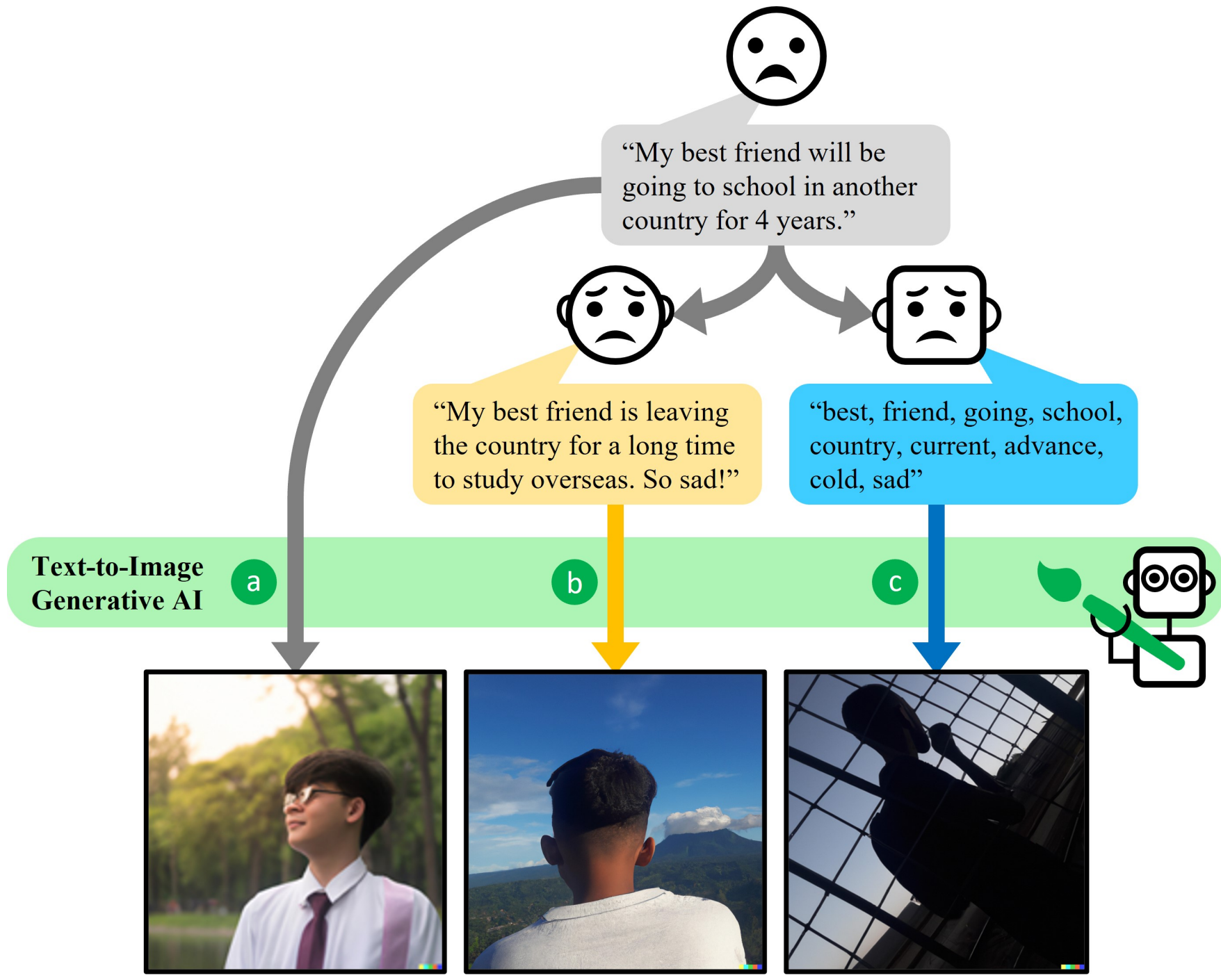

**Figure 1: Concept diagram of prompt engineering. (A) The text-to-image generative AI model can generate images with free-form text, but the generated image might not well match the expected emotion because the text prompt is not optimal to the AI model. (B) Lay users might revise the prompt with trial and error, which is inefficient. (C) In this paper, we aim to develop an automatic prompt engineering method that generates human-readable and effective prompt revision to improve the image output with text-to-image generative models. The images were generated by DALL·E 2.**

We first conducted an interview study to observe how laypersons perceived the AI-generated images and how they edited text prompts to improve AI outputs. Based on our findings, we further developed a computational approach with explainable AI (XAI) techniques to automatically edit the text to prompt the AI model for better image generations. Using AI to generate emotional images with texts has many potential applications, which we discuss at the end of this paper. Our **contributions** are:

- We reported the results of an interview study (n = 19) on how laypersons understand the text-to-image model and edit text prompts to improve the image output in the use case of emotion expression.
- We proposed RePrompt, an automatic explainable prompt-refining pipeline. Following our findings from the interview study, we selected intuitive text features and developed a proxy model to analyze the feature effects on the AI-generated image with a large dataset. Applying model explanations, we curated a rubric to allow automatic prompt-editing to improve the emotional expression of images generated by the AI model.
- We conducted a simulation study and a user study (n = 197) to evaluate our proposed method against other methods. The results suggested that our proposed method could significantly improve the emotional expression of the AI-generated images, especially for negative emotions.
- We end by discussing design implications, the generalization of our method, and potential applications.

## 2 BACKGROUND AND RELATED WORK

### 2.1 Text-to-Image Generative AI and Vision-Language Model

HCI researchers have been exploring how to harness cutting-edge AI technologies to support creativity, including crowd ideation [18, 83], music creation by novices [44], and news illustrations [43]. Text-to-image generative AI and its applications are outstanding and increasingly drawing attention from both academia and industry, due to the power of enabling laypersons to create visual art with plain language. Unlike prior tools providing guidance or stimuli during the creation process, state-of-the-art models (e.g., VQGAN-CLIP [19, 65] DALL·E 2 [55, 61], and Stable Diffusion [16, 66]) have shifted the creation process to the machine while leaving only the text prompt design to users. Due to the ambiguity of natural language and the imperfection of AI models, it remains essential to understand 1) how accurately generative models can capture user intent in text prompts and 2) how to engineer text prompts to optimize the output.

The "brain" in text-to-image model that learns the text-image relationship is the pretrained vision-language cross-modal model (e.g., CLIP [60] and ALIGN [31]). It learns representations of texts and images in one unified semantic embedding space. It calculates the cosine similarity of a pair of embeddings to determine their semantic closeness. With the feedback from the language-image model, generative models iteratively improve the output image by optimizing the semantic closeness between the image and text prompt. Hessel et al. introduced the CLIPScore and found that CLIP could be used for image captioning evaluation with good performance [29]. These empirical studies suggested that the CLIP-based generative models capture the semantic meanings in texts and images. CLIPasso utilized CLIP to measure how much a sketch matched a target image, which enabled a model to generate sketches with multiple levels of abstraction while maintaining the essential structure and essential visual components of the subject drawn [79].

Wang et al. applied CLIP to calculate the embedding similarity of image-label pairs (e.g., an image with the label "good photo" or "bad" photo) to assess both the quality perception (e.g., "high quality" vs. "low quality") and abstract perception (e.g., "complex" vs. "simple") of images [80]. Likewise, Bondielli and Passaro explored CLIP as a zero-shot classifier of image emotions [6]. Using CLIP and VQGAN trained on the WikiArt dataset [49], Galanos et al. generated emotional images with a constrained set of short prompts with four emotions (e.g., a happy cityscape) [22]. The results of their small study with 32 AI-generated images suggested that human raters could recognize the intended emotion behind the image. However, it is still unclear if the AI models could generate precise images with more complex and natural emotional text prompts (e.g., "My best friend will be going to school in another country for 4 years."). These studies showed the potential of using CLIP to assess emotion expression in images. In our work, we continue this line of research by developing a new prompt engineering method to augment longer emotional text expressions.

## 2.2 Prompt Engineering for Generative AI

Prompt engineering for large language models (LLM, e.g., GPT-3 [9]) and text-to-image models (e.g., DALL·E 2) refers to creating or adjusting prompts to improve the model output for tasks like natural language understanding [32], natural language generation [39], sentiment and factual knowledge elicitation [72], image recognition [91], and image generation [41]. Given a pre-trained generative model (e.g., GPT-3 and DALL·E 2), prompt engineering methods are alternatives to model fine-tuning, especially for cases where the model has a large number of parameters to tune [39] or when the full model is not accessible for parameter tuning (e.g., DALL·E 2).

Prompt engineering, also known as prompt tuning [38], prompting [39, 91], prompt-based learning [40], and prompt programming [62], has gained attention from both AI and HCI communities. Many prior works have focused on gradient-based prompt learning methods for automated prompt template curation. Shin et al. proposed AutoPrompt for masked language models (e.g., BERT [20]) to adjust initial prompt text by adding a set of learned texts as trigger tokens [72]. Instead of adjusting text prompts, Li and Liang developed Prefix-Tuning to prepend a learned task-specific vector to the original prompt vector for the table-to-text generation and text summarization tasks [39]. Likewise, Lester et al. showed that prompt-tuning is simple yet effective for specific downstream tasks (e.g., natural language understanding) [38]. Other works have explored gradient-based methods for vision-language tasks. For example, CoOp learned the initial text prompts' context as vectors to fill a prompt template, which improved image recognition performance compared to directly using CLIP embeddings [91].

These gradient-based automated methods, however, were not human-readable, which might hinder applications in human-AI collaboration [84]. Wu et al. proposed AI Chains to decompose prompts for complex tasks to improve system transparency, controllability, and task outcomes [84]. Using data visualization to promote understanding and debugging, Strobelt et al. proposed PromptIDE for exploring prompt variations and their corresponding outcomes [75]. Our work falls in this domain of interpretable prompt engineering by leveraging usable information from AI models [54, 73, 82, 85] and providing interpretable automated edits.

For the specific field of text-to-image generative models, prompt engineering research is nascent. The image generation task is different from the NLP tasks regarding the output and evaluation methods. We identified several pioneer works in this research line. Liu and Chiton explored short prompts with several pre-set keywords of subjects and styles and found that different ways of phrasing (e.g., "a painting of love in the abstract style" vs. "love abstract art") had no effect on the quality of AI generated images [41]. Oppenlaender analyzed practitioners' posts in online communities on text-based generative art prompts and identified five major types of prompt modifiers (i.e., subject terms, style modifiers, quality boosters, repetitions, and magic terms) to improve image generations [57]. However, these works were either heuristic-driven based on manual, post-hoc observation or not explicitly implementable. In this work, we propose a novel prompt engineering approach that is data-driven and interpretable for text-to-image generation with a focus on emotional expression.

## 2.3 Emotional Expression and Assessment in Visual Art

As an important means to communicate information and emotions of humans, visual art (e.g., painting and photography) is prevalent in our daily life and can now be created by AI (e.g., DALL·E 2 [55]). Although visual art evokes an emotional response, it remains difficult to understand the mechanism of the viewers' emotional response to an artwork [27, 70]. The subjective evaluation of an artwork depends on the artwork's form and expressions, the audience's understanding of the content and meanings, the audience's life experience related to the artworks' expression, and the audience's active attempt to process the stimulus [53]. Although image quality assessment in computer graphics and computer vision is a long-standing domain and many evaluation methods have been developed, quantifying emotional expression in images or visual art is still challenging. Psychologists have developed numerous instruments to evaluate the emotional expression in visual arts through audience's perception and reaction. Hagtvedt et al. developed a Likert scale questionnaire to measure the affective and cognitive components involved in the perception of visual art [27]. Likewise, Schindler et al. developed the Aesthetic Emotions Scale, which contains 21 subscales to assess the emotional signature of responses to artworks and emphasizing the perceived aesthetic appeal [70]. Besides self-report scales, researchers also assessed behavioral and physiological signals to establish objective measures, e.g., cardiac signature through electrocardiogram [76], facial electrocardiogram [23], and skin conductance [77]. However, due to needing specialized equipment, these objective measurements can be hardly deployed at scale. Self-report scales are therefore still commonly used in research as convenient measures.

The aforementioned measures are based on human perception and responses. Another line of research, which automatically evaluates image quality and aesthetics with machine learning, is attracting attention [26, 35, 50, 87]. Based on findings that humans share common emotion perception of certain images due to human nature and cultural background, researchers have curated image datasets to assess human emotions, e.g., the International Affective Picture System [37] and Open Affective Standardized Image Set [36]. Based on these works and increasingly available data, the advancement of pre-trained vision-language models (e.g., CLIP mentioned in Section 2.1) enables new tools to assess images at scale. Bondielli and Passaro [6] conducted a study suggesting that, using CLIP embedding, the zero-shot classification accuracy for eight emotion types on an emotional image dataset [88] was 0.49, which was much more accurate than using ImageNet CNN (accuracy = 0.28) and slightly more accurate than the result of fine-tuned CNN (accuracy = 0.48) [88]. Inspired by prior studies, we adapted both objective (i.e., CLIP embeddings) and subjective (i.e., human ratings) measures to our evaluation scenario (see details in Section 5 and 6).

## 2.4 Explainable AI and Applications in HCI

Explainable AI (XAI) techniques have become imperative due to the complexity of AI systems, and many studies have shown that XAI can increase user trust and understanding [4, 81]. Depending on the input data types and the machine learning models, numerous XAI techniques have been developed to explain model behaviors, e.g., LIME [63], SHAP [46], Anchors [64], and saliency maps [71]. For human-AI collaboration scenarios, XAI could also provide crucial support to bridge users and the AI system. In recent works, XAI has been used to support computer-aided translation [17], music creation [45], AI-directed crowd ideation [83], and concise feedback generation [82]. Our work contributes to this research line by applying XAI to understand how the features of prompts affect the output of the text-to-image generative models and develop interpretable prompt engineering to effectively improve emotional expressiveness of text-to-image generative models.

# 3 RESEARCH QUESTIONS AND METHODOLOGY

We first illustrate our research questions and general methodology before introducing our studies and technical method.

## 3.1 Research Questions

We are interested in 1) understanding how well the state-of-the-art text-to-image generative models could generate images with emotional texts as prompts, and 2) exploring how to improve the emotion expression of the images generated. For these overarching goals, we propose three research questions:

- RQ1. How would laypersons **perceive** the AI-generated images regarding emotional expressiveness and strategically **refine** the text prompts to improve image generation?
- RQ2. How can text prompts be **automatically refined** to generate better images regarding emotional expressiveness?
- RQ3. How **effective** is RePrompt at improving emotional expression in image generation?

## 3.2 Methodology

To answer RQ1, we first conducted interviews to assess how laypersons understood the text-to-image generative AI model (i.e., DALL-E 2) and how they developed prompt-editing strategies through trial-and-error with our prepared emotional texts (Section 4). We performed a thematic analysis of participants' utterances and identified their strategies for editing the text prompts. Participants developed strategies based on their intuition after a few attempts, which suggests the need for understanding the AI model and for effective support for prompt refinement.

To address RQ2, we developed RePrompt, an XAI-based automatic prompt engineering method. Its pipeline consisted of two major steps: 1) understanding what prompt features could lead to better output images by text-to-image generative models, and 2) automatically revising text prompts to achieve better output images. In Step 1, we selected word-level text features (e.g., number of nouns) inspired by our interview study findings and trained machine learning models to predict the quality of AI-generated images. We applied XAI techniques to understand feature contributions to image quality. Then, we curated a rubric to algorithmically revise a given prompt by manipulating word-level features (e.g., adding or removing words), which is easy to understand. We describe the technical details in Section 5.

To evaluate the effectiveness of RePrompt (RQ3), we conducted a simulation study with computational metrics and an evaluation user study to compare the quality of images generated from RePrompt text prompts and other text methods. Next, we introduce the text-to-image models and text dataset used for our user studies and the development of our technical prompt engineering approach.

## 3.3 Text-to-Image Models

**VQGAN-CLIP** [19] is a popular open-source generative model, which has been used to investigate prompt engineering methods [41–43]. We chose this model to generate 10,000 images to train our proxy model, due to its openness and flexibility. VQGAN-CLIP leverages VQGAN [21] to generate images and CLIP [60] to guide the generative model. We used the VQGAN version pretrained on ImageNet with the 16384 codebook size and the CLIP version with the ViT-B/16 vision transformer [60]. Following Liu and Chilton [41], we set the resolution at 256 × 256 for image generation and limited the maximum number of iterations to 300 steps.

**DALL·E 2** [61] is a state-of-the-art generative model. Instead of using VQGAN, DALL·E 2 adopts a diffusion model to generate the image while still using CLIP to assess image-text alignment and guide image generation. From our pilot testing, DALL·E 2 could generate much more photorealistic images than VQGAN-CLIP, despite using art style keywords in prompts. Another difference between VQGAN-CLIP and DALL·E 2 is that DALL·E 2 has filtering mechanisms to mitigate the generation of deceptive and harmful content [56], while VQGAN-CLIP does not. For our user studies, we manually screened out improper texts to avoid unexpected emotional experience of our participants. It is available only via the official website and no API is available yet. Therefore, we manually input the text prompts and download the generated images. It took around 10 seconds for DALL·E 2 to generate a set of four high-resolution images for a given text prompt. DALL·E 2 is also faster than VQGAN-CLIP at producing images with the same resolution, which is necessary for real-time user interaction.

## 3.4 Emotional Text Dataset

**EmpatheticDialogues** [90] is a frequently used text dataset for training empathetic AI chatbots [59], which fits the need in terms of our research questions. It consists of 25k conversations grounded in emotional situations with emotion labels. For example, for an emotion label "proud", one emotional situation is "I finally got that promotion at work! I have tried so hard for so long to get it!", followed by a dialog around this situation. We used only **emotion labels** and **situation texts** in our studies, not the following dialog. As we introduced in Section 1, the focused scenario in this work is peoples' daily practice of emotion expression, e.g., sharing via messages to friends or posting on social media. Therefore, the situation texts in the EmpatheticDialogues dataset fit our scenario as they are natural emotion expressions grounded in people's real emotion experiences and contexts. Another reason for choosing this dataset was its high text quality. According to our sanity check, this dataset is clean of typos, random symbols, and odd acronyms that are found in some Twitter datasets. We randomly selected a small subset for the pre-defined text prompts used in our interview study (see Section 4.1). Later, we used a larger subset from this dataset as text prompts for developing our RePrompt method (see Section 5).

# 4 INTERVIEW STUDY

To answer our first research question, we first conducted an online interview study by using the think-aloud method.

## 4.1 Emotional Texts

We randomly selected 200 emotional texts with 10 emotions: *joyful*, *sad*, *angry*, *afraid*, *lonely*, *excited*, *proud*, *surprised*, *trusting*, and *anxious* (20 texts/emotion). We excluded instances with *trusting* as the emotion due to low text-emotion alignment in the randomly selected text instances. For example, the text "I knew I shouldn't have trusted my brother with my dog" does not match the emotion label (i.e., *trusting*). We also dropped the *anxious* emotion because many words (e.g., "suffering") were forbidden due to DALL-E 2's content policy. For the emotion of *afraid*, *lonely*, and *sad*, we excluded 14 texts due to strongly negative words (e.g., "die"). The final set consisted of 146 texts and their corresponding images.

## 4.2 Procedure

After receiving the participants' informed consent, the experimenter first introduced the study purpose of exploring how good could AI generate images according to text prompts and their emotional expression. The experimenter shared their screen of the OpenAI DALL·E 2 website (https://labs.openai.com) with the participant over Zoom and pasted the prepared text into the input box on the website. Then, the experimenter guided the participant through the following steps for 10 rounds:

1. Read and verbally rate (from 0 to 100) the text-emotion alignment (i.e., "how much do you think this text expresses the [emotion label, e.g., lonely] emotion?").
2. View the AI-generated images (four images per prompt) according to the text prompt.
3. Verbally rate the image-text alignment (ITA; i.e., "how much do you think the images express the text?") and image-emotion alignment (IEA; i.e., "how much do you think the images express the [emotion label, e.g., lonely] emotion?").
4. Edit the text and repeat steps 1–3.

For the text editing, we sent the original text to the participant and received the edited text via the chat function in Zoom. We emphasized to the participant that they could freely edit the text with the purpose of prompting the AI to generate better images for emotional expression while trying to maintain the context of the original text. We asked the participant to explain their strategies for editing the text during or after the editing.

The rating process was to help the participant to think through how much the images expressed the content of the text prompts and understand how the AI-generated images matched the text. We did not use the ratings for image evaluation due to the possibility of cognitive bias (i.e., the IKEA effect [52]). Instead, we evaluated the images with objective measures and external validator ratings (in Section 6). Each participant completed 10 rounds. For the first two rounds, participants iterated the editing (steps 2 to 4) two times to help them better practice and foster their editing strategies. At the end of each interview, we asked the participant to reflect on their understanding of AI and their text-editing strategies.

## 4.3 Participants and Data Collection

We recruited 19 participants from a university mailing list. They were 8 males and 11 females, with ages ranging from 19–39 years old. All were students with diverse majors, e.g., engineering, business, psychology, computer science, pharmacy, and social science. The experiment took 40–60 minutes and participants were compensated $10 SGD ($7.20 USD). We conducted the study online via a Zoom audio call with screen recording, to which the participant consented.

We performed a thematic analysis of participants' utterances and their edited texts using open coding [25]. Thematic coding was performed by one co-author with regular discussion with a senior co-author.

**Table 1: Participants' editing strategies with examples.**

| Strategy | Original Text | Edited Text |
|---|---|---|
| Emphasizing emotion by adding or adjusting emotion words | My best friend and I had been roommates all through college. We each moved out on our own over the summer. I miss her. | My best friend and I had been roommates all through college. We each moved out on our own over the summer. I miss her. I feel lonely. |
| | My friend's girlfriend cheated on him. I've never seen him so destroyed. | My friend's girlfriend cheated on him. I feel like he is very disappointed and confused. |
| Emphasizing emotion by adding context | I was scared when I had to go home alone the other day. | I felt very scared when I had to walk home alone at quiet and dark night. |
| | I finished college last May after 4 long years. | I finally finished college after working hard for my degree for 4 long years. |
| Simplifying text by deleting sentences or paraphrasing | Went out with my friends last night after not seeing each other for a year. I was in another country for work and just got back. | I went out with my friends to party. |
| | I was mad when my boss yelled at me. I guess I was late | I am angry at my boss. |
| Increasing concreteness | I was shocked when I was able to pass my history class so easily when I didn't study much ever. | I was so surprised for passing the history class, I only made so less note and didn't prepare well while the teacher still got me passed. |
| | Sometimes I want privacy, but now that I have it, I miss my roomies. | Now that I have my own room, I realise how much I miss my roommates. |
| Changing self-expression (1st-person view) to image description (3rd-person view) | I was so proud of my dad when he retired. He deserved it. | smiling daughter handing a "Happy Retirement" card to smiling old dad with white hair. |
| | I was all alone the other day when my kids left | devastated man alone in room |

## 4.4 Results

All participants used trial-and-error initially but updated their editing strategies as they gained understanding of the AI model (DALL·E 2).

*4.4.1 Participants' Understandings of the Text-to-Image Model.* We identified five themes from participants' understandings of DALL·E 2 from the interviews.

**The AI could understand simple and concrete texts, instead of complex and abstract texts**. Participants found that the AI model ignored some context when the text consisted of many subjects or objects in the text. Sometimes the AI failed by just showing some gibberish in the output image. P6 commented that there were "*a lot of things for the AI to consider so he had to stop,*" and P17 thought that "*maybe the sentence is too long and too complex*" when seeing unexpected images.

**The AI did not understand some specific concepts.** Participants noticed that the AI did not understand concepts like movie names (e.g., the horror movie "IT") and commodity names (e.g., the game console "Nintendo switch"). In these cases, DALL·E 2 failed to portray relevant elements in the output image. P16 assumed that "*maybe there's no data about Nintendo switch in the AI.*"

**An objective description of the expected image seems better for AI to understand.** P14 asked the experimenter if he was allowed to just use keywords to describe the imagined image in his head and found that doing so worked quite well for some instances, commenting "*I think you have to be quite objective in a way you describe what's going on.*" P5 got a similar feeling and tried to directly describe the expected image according to the emotional text, as he mentioned "*imagine how you want to reach out to [object in the image].*"

**The AI was not quite accurate, and it was sometimes difficult to understand it when it failed to generate the expected images.** Many participants encountered some failed editing to improve the image output and found it difficult to understand when and why the AI failed. For example, some of the comments included, "*I think, at this stage, AI couldn't accurately understand human language*" (P8), "*AI wouldn't really be accurate in that case*" (P12), and "*AI didn't follow my text, I . . . have no idea why that's happening in this case*" (P5).

The observations from participants' understanding of the AI not only confirmed some of our prior knowledge (e.g., text concreteness) but also demonstrated the AI's bias towards certain prompt styles. This helps us make sense of the participants' strategies. Additionally, the observations suggested that failure in trial-and-error leads to bad user experience, which warranted support to help users better interact with the AI system.

*4.4.2 Participants' Editing Strategies.* Most participants tried to find a simple and unified strategy through iterations. After being successful for one instance, they were likely to follow the same strategy until it failed. Overall, the participants were engaged in the interview and provided a variety of inspirations for our design of RePrompt. We list the top 5 strategies with examples in Table 1.

Overall, we abstracted participants’ strategies into two principles: 1) make the emotion and key concepts in the context concise and straightforward, and 2) make the elements and description concrete. However, participants developed strategies based on their intuitions, with a few attempts lacking precision and global understanding of the AI model. It remained unclear whether the strategies could generally improve image generation and how to automate effective strategies in a principled way. Hence, we needed a data-driven method to automatically refine text prompts.

## 5 TECHNICAL METHOD OF REPROMPT

Through our interviews, we learned how laypersons attempted to edit text prompts to improve the emotional expressiveness of AI-generated images. In this section, we introduce RePrompt, a novel prompt engineering method to automatically edit text prompts, answering our second research question:

RQ2. How to **automatically refine** text prompts to generate better images regarding emotional expressiveness?

### 5.1 Feature Curation

To automatically refine text prompts, we first need to understand which and how context features in the text prompts influence the quality of the image generated by the AI model. Our feature curation process was informed by our interview study results, prior works of understanding large language models (e.g., [54]), and text-based emotion analysis with machine learning [1, 59]. We noticed that many sophisticated context features have been deployed for emotion analysis [1, 59], such as n-grams, word embedding, and part-of-speech, but not all of them are suitable for our scenario. Considering that our goal is to develop human-interpretable prompt engineering methods, we chose intuitive and easy-to-adjust features (e.g., the number of nouns in a text). We omitted structural features (e.g., constituent parse trees), since laypersons do not have the linguistic training to appreciate them easily.

We designed the features as follows: 1) identify the part-of-speech (POS) of each word in the text prompt, and 2) count occurrences of each POS type (e.g., number of nouns), and compute the mean concreteness score of words in each type of POS (e.g., mean concreteness of adjectives) as features. These word-level features meet our requirements of being easy to understand and adjust. Importantly, they allow us to relate the data-driven results based on these features to the findings from our interview study. The features of word counts reflect the text complexity while the features of word concreteness indicate the text concreteness. We obtained word concreteness from the English word concreteness dataset [10], which contains human ratings of word concreteness for 40,000 generally known English word lemmas. We considered 20 features in total, as shown in Figure 10 in the Appendix. These features are tunable as we can delete, add, or replace words to change the feature values. For example, one feature is the number of nouns in the text, and we can delete or add nouns to easily adjust the feature value. Given the curated features, we then need a model to link the features to the quality of AI-generated images.

### 5.2 CLIP Score for Image Quality Assessment

We adopted the CLIP Score for the image-emotion alignment (IEA) and image-text alignment (ITA) as measures of image quality. Note that the image quality here is not about aesthetic quality, but semantic alignment to the text prompts. As mentioned in Section 2.1, CLIP is a pre-trained vision-language model that can represent both text and images in one unified semantic embedding space [60]. The cosine similarity of a pair of text and image embeddings represents their semantic closeness. CLIP was trained on a dataset of 400 million image-text pairs collected from the Internet [60]. It is reasonable to assume that the training data contain myriad concepts of both concrete objects and abstract descriptions. Besides using CLIP for text-to-image generation models (e.g., DALL·E 2 and VQGAN-CLIP), researchers have also explored the power of CLIP for many other downstream tasks such as zero-shot image captioning evaluation [29] and image property assessment [80]. Highly relevant to our application of CLIP, Bondielli and Passaro’s study suggested that CLIP could enable zero-shot image emotion recognition with a similar accuracy compared to the fine-tuned CNN classifier based on a pre-trained ImageNet model [6, 88]. Inspired by these works, we applied CLIP to calculate the IEA and ITA in our study. The CLIP Score is defined as the cosine similarity of the resultant CLIP embeddings (i.e., **c** and **v**) of the interested text-image pair (Equation 1) [60]. For IEA, the two embeddings correspond to the image and the emotion label (e.g., “sad”); for ITA, they correspond to the image and the text (e.g., “My best friend will be going to school in another country for 4 years.”). We used the ViT-B/32 version of CLIP [60].

$$\text{CLIP Score}\,(\mathbf{c},\ \mathbf{v}) \ = \ cos\,(\mathbf{c},\ \mathbf{v}) \tag{1}$$

We note that CLIP could not identify meaningless input (e.g., “#%@)%I”); these symbols led to meaningless CLIP Scores. For example, the CLIP Score of (“#%@)%I” ,“dog”) is higher than the CLIP Score of (“animal”, “dog”). The problem of junk characters can be addressed by employing spell check. We checked the used text dataset in our study and found no such junk input. We list examples of CLIP Scores for IEA and ITA, as well as cautions for using the CLIP Score for text with junk input, in Table 3 in the Appendix.

### 5.3 Proxy Model and Feature Analysis

Given the CLIP Score as the computational measure of image quality, we developed machine learning models to predict the image quality with the curated features in Section 5.1. The first goal is to understand how these features affect image generation. One effective way for such feature analysis is using explainable AI (XAI) techniques (e.g., SHAP [46]). However, computing model explanations for feature analysis requires calling the model many times, which is unscalable for generative AI models. To solve this problem, we introduce a proxy model to estimate the image quality score instead.

To train the proxy model for predicting image quality (Figure 2, Part 2), we first need enough training data. We used 10,000 emotional texts with 32 emotions from the EmpatheticDialogues dataset (Section 3.4) and generated one image for each text instance with the VQGAN-CLIP model and calculated the CLIP scores (i.e., IEA and ITA) (see Figure 2, part 1). We gradually increased the text sample size starting from 5,000 texts and found that the model training saturated with robust results after 8,000 texts. Note that we did not use DALL·E 2 to generate these images since we did not have API access. We will show later that using VQGAN-CLIP to generate data for training the proxy models did not affect the evaluation results. We used the same VQGAN-CLIP configuration as [41] and generated images on a local workstation server with an NVIDIA GeForce RTX 3090 GPU.

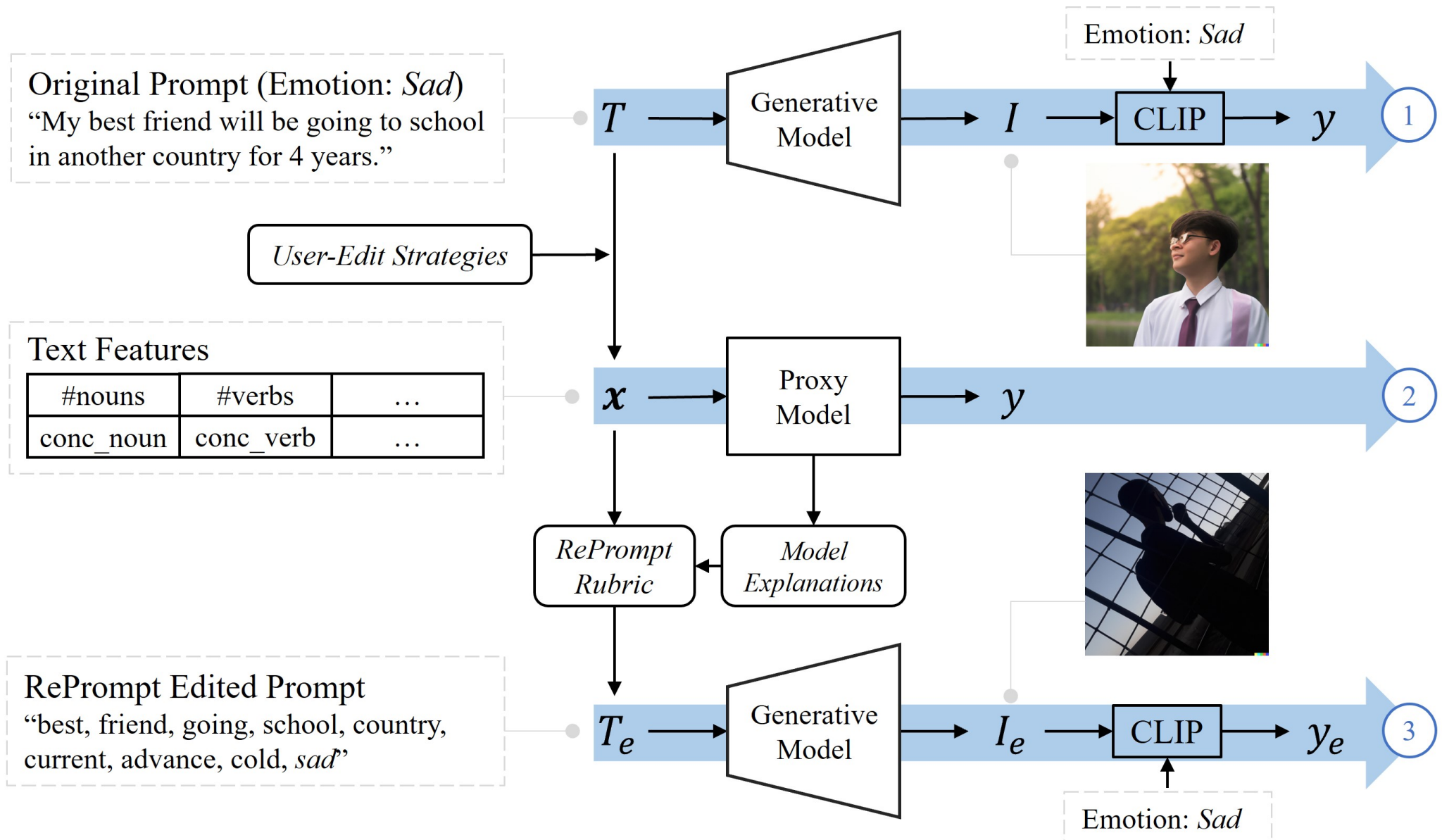

**Figure 2: RePrompt pipeline. Part 1 (the upper flow) is the process of generating the image ($I$) with the original text prompt ($T$) by the generative model and computing the CLIP Score of image-text alignment ($y$). Part 2 (the middle flow) is the process of training the proxy model with text features ($x$) to predict $y$. "#nouns" is the number of nouns, and "conc_noun" is mean concreteness of nouns. Part 3 (the lower flow) is the process of generating the new image ($I_e$) with the edited text prompt ($T_e$) by the generative model and computing the new CLIP Score of image-text alignment ($y_e$). $y_e$ is expected to be higher than $y$. Text feature curation from $T$ to $x$ was inspired by the participants' prompt-editing strategies in our interview study. Model explanations and the following RePrompt rubric curation, which are based on the trained proxy model in Part 2, are key for automated prompt-editing.**

We modeled the score prediction as a classification problem by binarizing the alignment scores (>mean or not), which made the following model explanations more intuitive to understand compared to regression models [83]. With the text features and classification labels, we fitted several machine learning models (Random Forest [7], XGBoost [14], LightGBM [34], and Multi-layer perceptron) to separately predict IEA and ITA, and found that LightGBM achieved the best performance with 5-fold cross-validation (AUC = 0.60 for IEA, AUC = 0.73 for ITA). Since the primary target was to improve the emotional expressiveness in the generated images, we focused on the IEA prediction model to investigate model explanations, which were used to derive the rubric for automatic text editing.

## 5.4 Feature Analysis by Model Explanations

We employed 1) SHAP explanations to see which features were important and 2) Partial Dependence Plots (PDPs) to identify the optimal feature value ranges for high predictions. For the model explanation of feature contribution, SHAP (SHapley Additive exPlanations) [46] is a popular XAI technique. SHAP computes the importance of each feature by ablating (removing) them from the model and noting the decrease in model performance. It uses the notion of Shapley values from game theory by permuting combinations of when to ablate, since order affects the results. SHAP provides one feature importance explanation for each instance, which can be aggregated to show an overall global explanation. A partial dependence plot of the salient features could additionally provide the distribution of feature contribution over the value range. Based on the identified feature value ranges, we curated corresponding rules for feature value tuning, as shown in Figure 3.

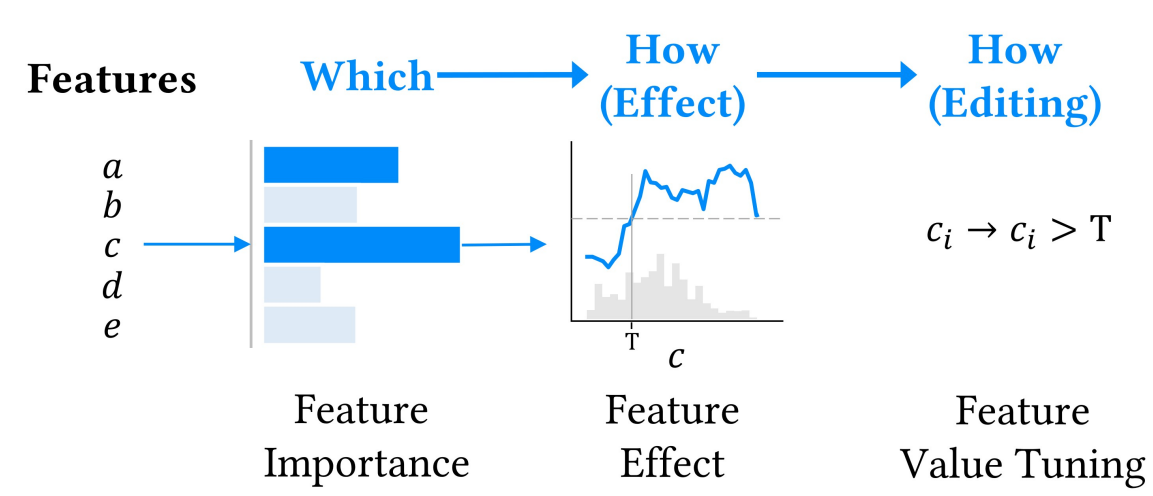

**Figure 3: Concept illustration of applying model explanations for feature analysis and feature value tuning. Given a machine learning model, we first calculated the global feature importance with SHAP values and selected the most important ones. Then, we utilized the partial dependence plot to analyze the feature's effect on the model output over the feature value distribution. By identifying the optimal feature value range, we finally curated the rule of feature value tuning.**

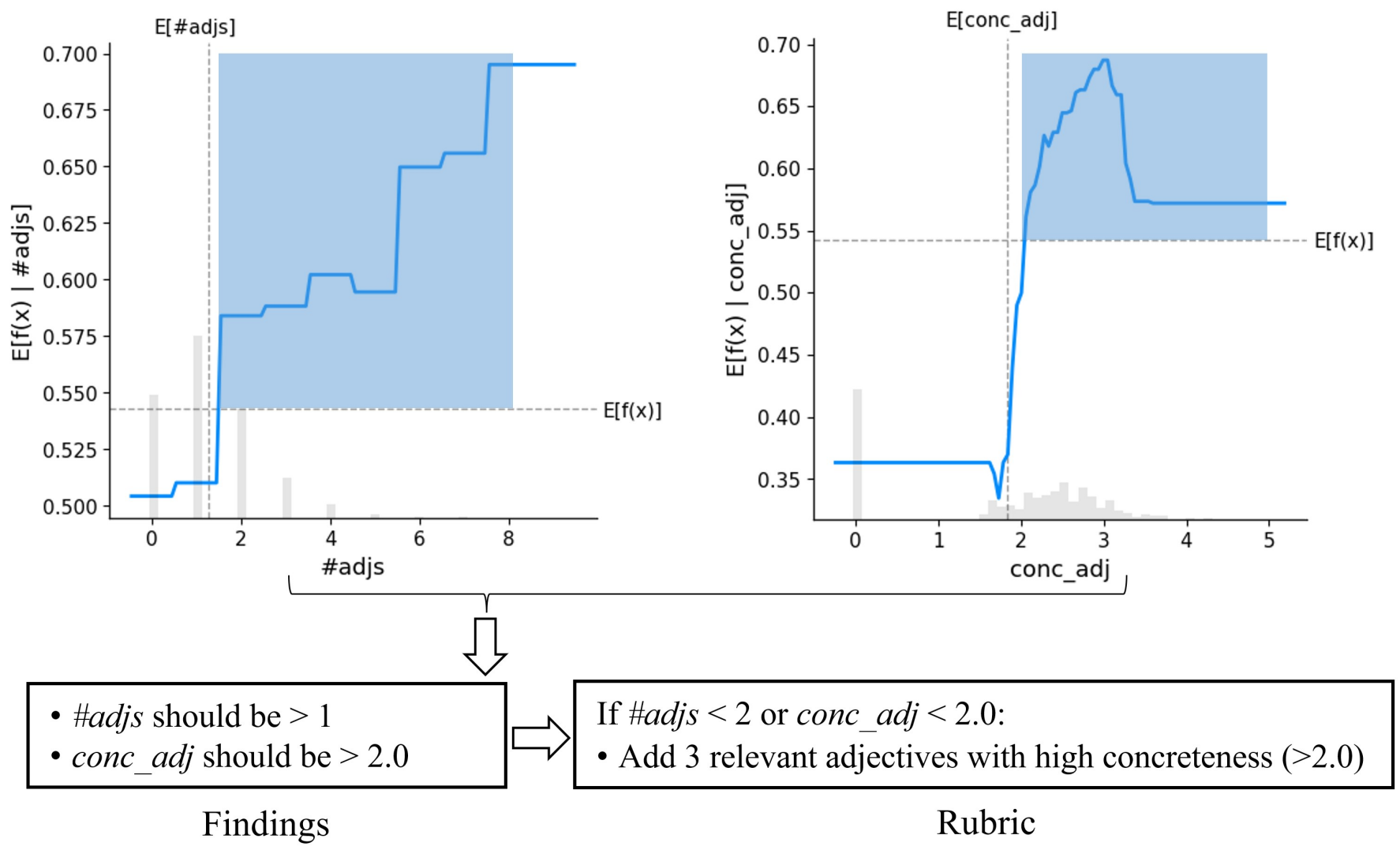

**Figure 4: Rubric curation for adjectives.**

**Selecting Salient Features**. We first applied SHAP to the proxy model to obtain the global feature importance for IEA prediction (see Figure 10 in the Appendix for the feature importance ranking). This helped to identify salient features that we should focus on. Note that we selected features according to not only the feature importance but also the ease of tuning values. Using an ablation method, O'Connor and Andreas found that nouns and verbs are more important than adjectives and function words in prompts for large language models regarding information loss [54]. Our results, with a different method, suggested that adjectives are as important as nouns and verbs. We eventually selected six features for this study, including the number of nouns (*#nouns*), the number of adjectives (*#adjs*), the number of verbs (*#verbs*), mean concreteness of nouns (*conc_noun*), mean concreteness of adjectives (*conc_adj*), and mean concreteness of verbs (*conc_verb*).

**Identifying Feature Value Ranges**. The partial dependence plots (PDP) illustrate the marginal model output over the value distribution of the target feature. We show the PDPs of two features of adjectives, *#adjs* and *conc_adj* in Figure 4 to demonstrate the process of rubric curation. We first identify the feature value range ($x$) according to Equation 2, where $f(x)$ is the model output (i.e., the probability of predicting High IEA by our proxy model).

$$\arg_{x} \left( f\left(x\right) > E\left[f\left(x\right)\right] \right) \tag{2}$$

The feature value ranges are highlighted with blue blocks in Figure 4. According to the findings of the targeted feature value ranges, we finally curated the rubric for text feature editing. The findings of optimal feature values are 1) the number of adjectives in the text should be larger than 1, and 2) the mean concreteness of adjectives should be larger than 2.0. These findings aligned with our interview study results and provided quantitative evidence with actionable implications for prompt refinement. To obtain the optimal feature values, we adopted a simple method of adding relevant adjectives with high concreteness scores. The rubric curation processes for nouns and verbs were similar, which we detail in Figure 11 and Figure 12 in the Appendix.

## 5.5 Rubric for Automatic Prompt-Editing

We list the rubric in Table 2. The rubric is based on our findings of feature analysis and satisfies three principles: be concise regarding key elements in images, be concrete regarding element descriptions, and be consistent with the key context of the original text. Note that for nouns and verbs, we adjusted only their numbers but not word concreteness, because adjusting word concreteness requires replacing original nouns and verbs, which might break the contextual meaning in the description. For the case of adding adjectives, we limited the selection to three relevant words from ConceptNet [74], since we found that adding more could not benefit performance, according to our test with 1-5 words.

Figure 5 presents one example that illustrates the prompt editing process of RePrompt. A) Given a text, we first labeled the part-of-speech (POS) of each word and discarded words that are not nouns, verbs, or adjectives. We then calculate the word saliency using the CLIP Score of the word (e.g., friend) and the full text with the emotion label appended (i.e., "My best friend will be going to school in another country for 4 years. Sad." in this example). We used the word-saliency ranking to determine which word to remove and which related word to add. In this example, because *#nouns* is 4, we removed the least salient noun (i.e., "years") according to the saliency order. B) We retrieve relevant words of the top-3 salient words (i.e., "friend", "going", and "school") from ConceptNet [74], following the same setting in related work [83]. We kept only adjectives from the retrieved words according to the rubric. C) For the retrieved words, we calculated the word saliency (same method as in step A) and looked up the word concreteness, then kept the three most salient words with required concreteness (>2.0). D) We appended the emotion label and finalized the output of the prompt revision. To summarize this example, we kept all adjectives and verbs, removed one noun (because *#nouns* >3), added three adjectives (because *#adjs* <2), and appended the emotion label.

**Table 2: RePrompt rubric for automatic prompt-editing consisting of three rules based on six salient and tunable features according to feature analysis with SHAP explanations.**

| Feature Type | Findings | Rules |
|---|---|---|
| noun | *#nouns* should be <4<br>*conc_noun* should be 3.5–4.2 | **if** *#nouns* >3 **then**<br>reduce nouns according to word saliency until *#nouns* = 3 |
| adjective | *#adjs* should be >1<br>*conc_adj* should be >2.0 | **if** *#adjs* <2 *or conc_adj* <2.0 **then**<br>add 3 relevant adjectives with high concreteness (>2.0) |
| verb | *#verbs* should be <3<br>*conc_verb* should be >2.0 | **if** *#verbs* >2 **then**<br>reduce nouns according to word saliency until *#verbs* = 2 |

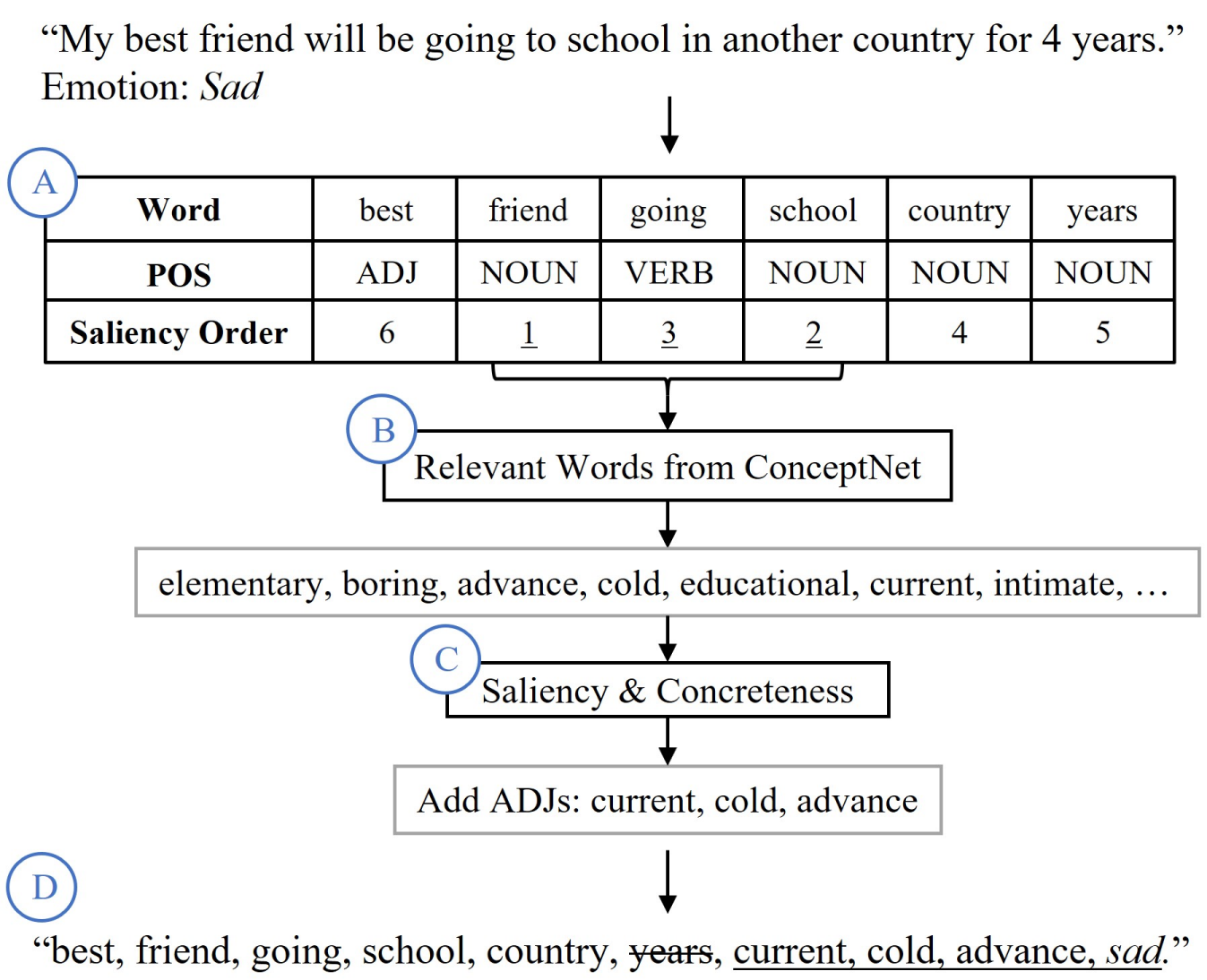

**Figure 5: An example of prompt editing with RePrompt.**

# 6 EVALUATION

After demonstrating our RePrompt method, we sought to answer our third research question:

RQ3. How **effective** is RePrompt at improving emotional expression in image generation?

## 6.1 Metrics and Conditions

To evaluate the efficacy of RePrompt, we conducted a simulation study and an evaluation user study. In the simulation study, we used CLIP Score to objectively measure the image-emotion alignment (IEA) and image-text alignment (ITA), following the same definition in Section 5.2. In the user evaluation study, we measured external validators' perception of IEA and ITA across images from different conditions.

We evaluated RePrompt against three other Prompt Editing Methods. Besides *Original Prompt*, *Manually Edited Prompt*, and *Automatic RePrompt*, we added a naïve automatic text-editing method, *Label Appended Prompt*, which simply appends the emotion label to the original text. Using the same text instances as in the interview study, we generated images using DALL·E 2. Figure 6 shows examples of texts and images across the four conditions.

## 6.2 Statistical Methods

We used linear mixed effects regression (LMER) models on various dependent variables, performed ANOVAs on the fixed main and interaction effects and post-hoc contrast tests for the specific differences identified. Since LMER models can accommodate missing data, we could analyze our experimental results with partially within-subjects independent variables. Due to the large number of comparisons in our analysis, we considered differences with $p < .001$ as significant and $p < .005$ as marginally significant. This is stricter than a Bonferroni correction for 50 comparisons (significance level = .05/50).

## 6.3 Simulation Study

To analyze objective image-emotion alignment and image-text alignment, we calculated the CLIP Score and fitted LMER models with the Prompt Editing Method as the fixed main effect and prompt ID as a random effect. Table 4 in the Appendix summarizes the goodness-of-fit of the LMER models and statistical significance of the ANOVA tests and Figure 7 illustrates the results.

**Image-Emotion Alignment (IEA) CLIP Score**. All three prompt editing methods resulted in images with significantly higher image-emotion alignment (IEA) scores compared to *Original Prompt* (Figure 7a). More importantly, images from *RePrompt* got significantly higher scores than images of other prompts. By contrast, *Manual Edited* got lower scores than *Label Appended*.

**Image-Text Alignment (ITA) CLIP Score**. *Manual Edited* and *RePrompt* had significantly lower image-text alignment (ITA) scores compared to *Original*, suggesting that these prompts changed the original text significantly. However, *Label Appended* had a significantly higher ITA score compared to *Original* (Figure 7b), suggesting that emotion labels are very related to the original text.

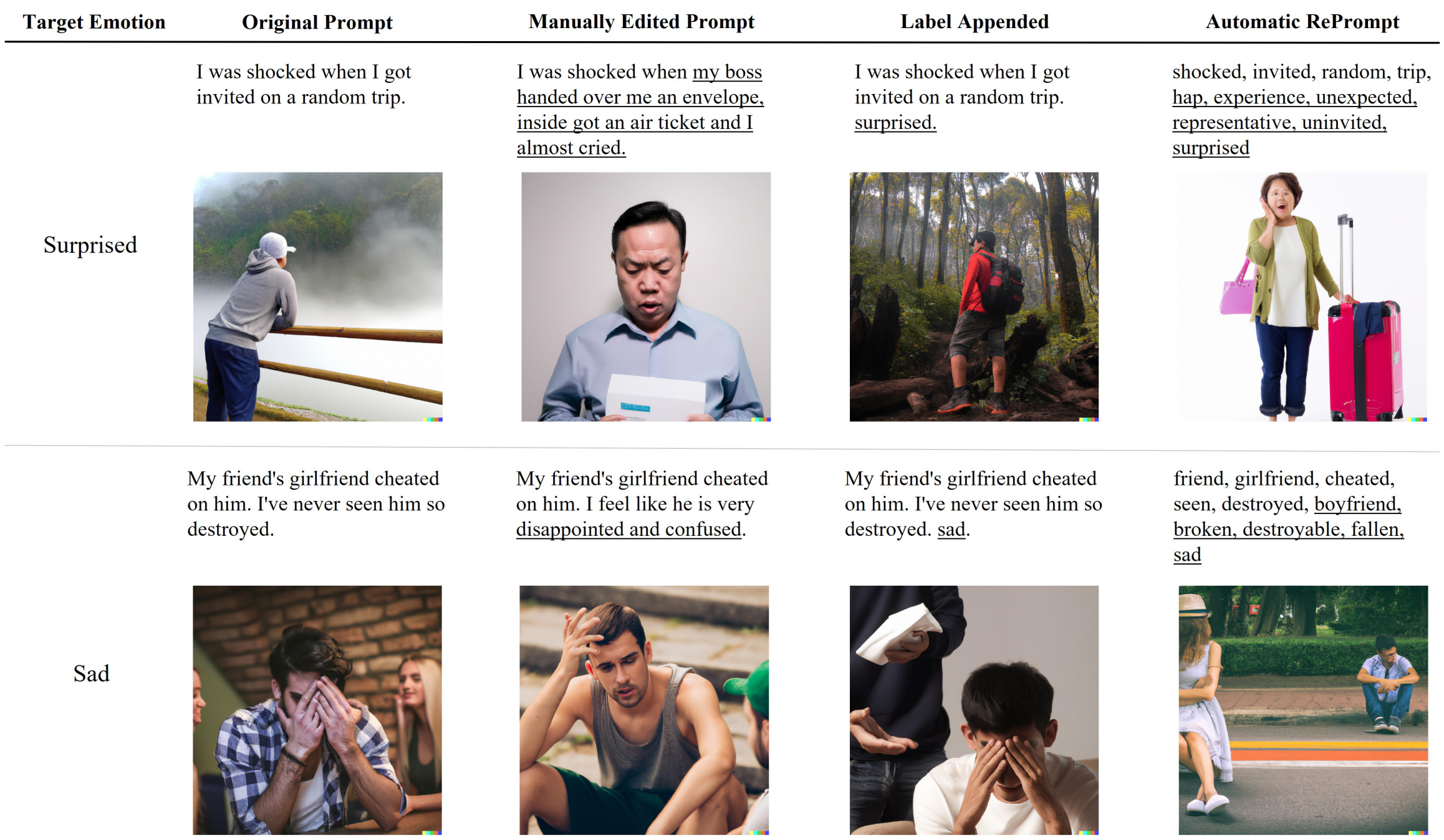

**Figure 6: Example prompts and AI-generated images across four conditions. The underlined portions indicate the added parts in the edited texts compared with the original texts. The images were generated by DALL·E 2.**

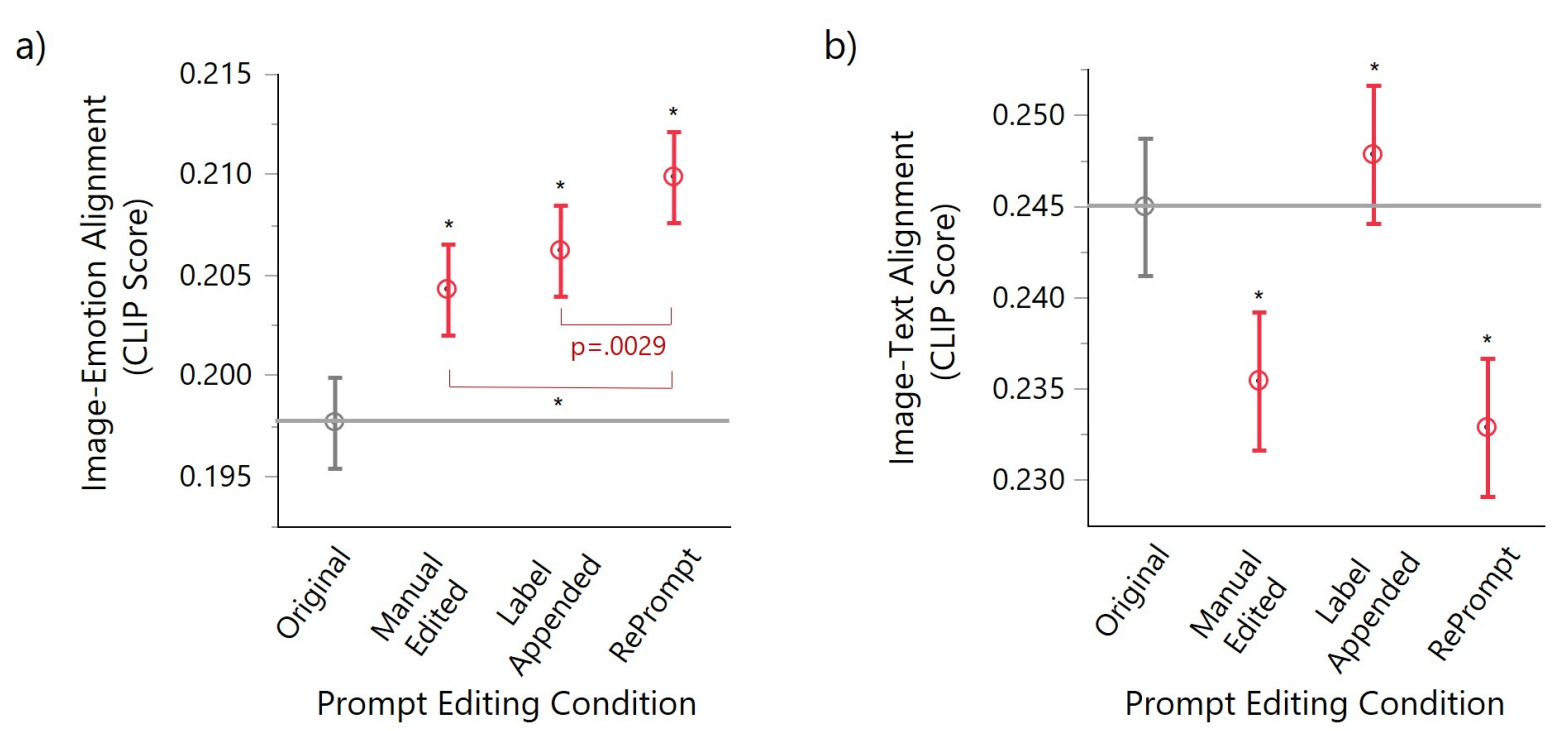

**Figure 7: Simulation results of CLIP Scores of image-emotion alignment and image-text alignment. * indicates significant difference at p < .0001 compared to the *Original* condition.**

## 6.4 Image Evaluation User Study

The simulation study measured image quality with computational metrics (i.e., CLIP Score), but people are subjective and diverse in perceiving emotions in images. Thus, it is necessary to evaluate with a user study. We invited a new set of participants to validate the image quality regarding IEA and ITA. We conducted a mixed-design experiment with Prompt Editing Condition (4 levels: Original Prompt, User Edited Prompt, Label Appended, RePrompt) as the independent variable. Since DALL·E 2 generated four images per text prompt, we randomly selected one for each instance. Eventually, we prepared 146 groups of images and each image group contained one image from each condition. We listed the screenshots of the full survey in Figures 13 - 26 in the Appendix.

*6.4.1 Procedure.* Participants were asked to evaluate emotional texts and corresponding images. The experimental procedure was as follows:

1. Introduction to experiment objective and consent to the study.
2. Screening quiz with a 4-item word association test [13] to assess English language skills and a 5-item emotional intelligence test (adapted from the Situational Test of Emotional

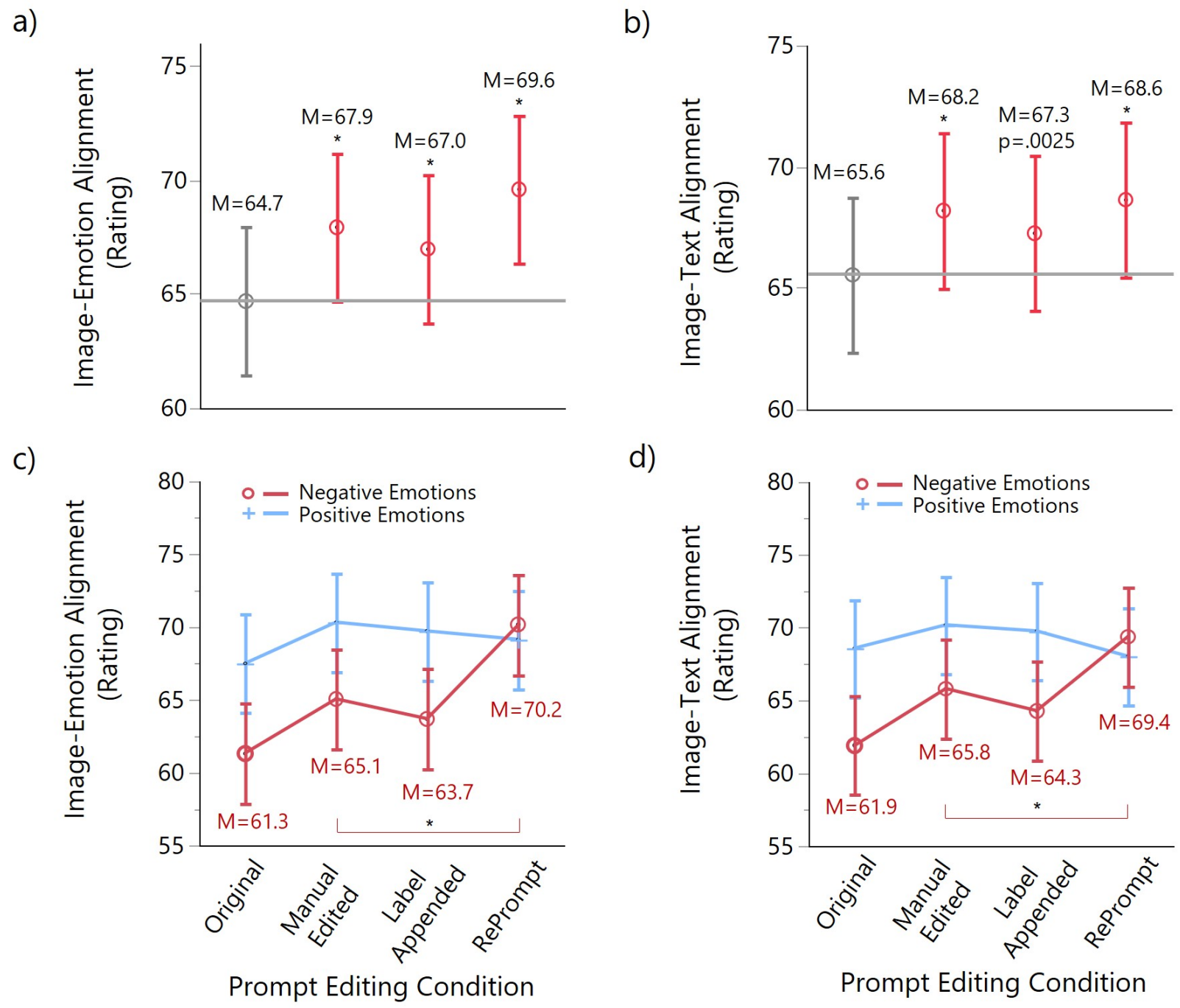

**Figure 8: Results of user ratings of image-emotion alignment and image-text alignment. P-values indicate difference from the Original condition if not specified. * indicates p < .0001.**

Understanding [47]). All answers must be correct to continue.

3. Assess 15 texts (randomly chosen) and corresponding image groups. For each message,
   a. Read text and rate on a 0–100 slider scale regarding how much the text (e.g., "my girlfriend dumped me the other day") expresses the labeled emotion (e.g., sad).
   b. Review four images (each generated from different conditions) and rank them regarding image-emotion alignment (i.e., how much each image expresses the labeled emotion in the text).
   c. Rate each of the four images on a 0–100 slider scale regarding image-emotion alignment.
   d. Write their rationale for their ratings. This was asked only three times to mitigate response fatigue.
   e. Review the four images again and rank them regarding image-text alignment (i.e., how much each image expresses the scene and elements in the text).
   f. Rate each of the four images on a 0–100 slider scale regarding image-text alignment.
   g. Write their rationale for their ratings. This was asked only three times to mitigate response fatigue.
4. Post-questionnaire on demographics.

*6.4.2 Participant and Data Collection.* We recruited participants from Amazon's mechanical Turk (criteria for participation: ≥5000 completed HITs, >97% approval). Out of the 721 workers who responded, 197 passed the screening and completed the survey (27.3% pass rate). Participants were 48.2% female, 19–72 years old (M = 40.5), completed the survey in about 36.9 min (median), and were compensated US$5.00. In total, 146 image groups were rated 2955 times (M = 20.24 times/group). The average aggregate-judge correlations [12, 83] were .311, .394, and .380 for text-emotion alignment, image-emotion alignment, and image-text alignment ratings, respectively.

*6.4.3 Results of Ratings.* We first fit LMER models with Prompt Editing Condition as a fixed main effect and participant as a random effect. Table 6 in the Appendix summarizes the LMER models and statistical significance of the ANOVA tests for the results of ratings and Figure 8 shows the results.

**Image-Emotion Alignment (IEA) rating**. Consistent with the simulation result, images in all three prompt editing methods resulted in significantly higher IEA ratings compared to the images with original texts, as shown in Figure 8 (a). Furthermore, RePrompt images were still significantly better than other prompt-editing methods. However, the practical differences between the RePrompt condition and other conditions were very small.

**Image-Text Alignment (ITA) rating**. Surprisingly, images in all the prompt editing conditions also had significantly higher ITA ratings compared to the original texts, as shown in Figure 8 (b).

**Confounder analysis.** We explored if we missed modeling important factors, and found that emotion type (i.e., positive emotion and negative emotion) had a strong effect. We adjusted our LMER models by adding Emotion Type and its interaction with Prompt Editing Condition as the fixed main effect. Table 7 in the Appendix summarizes the statistical modeling. As shown in Figure 8 (c, d), the interaction effect was significant statistically and practically.

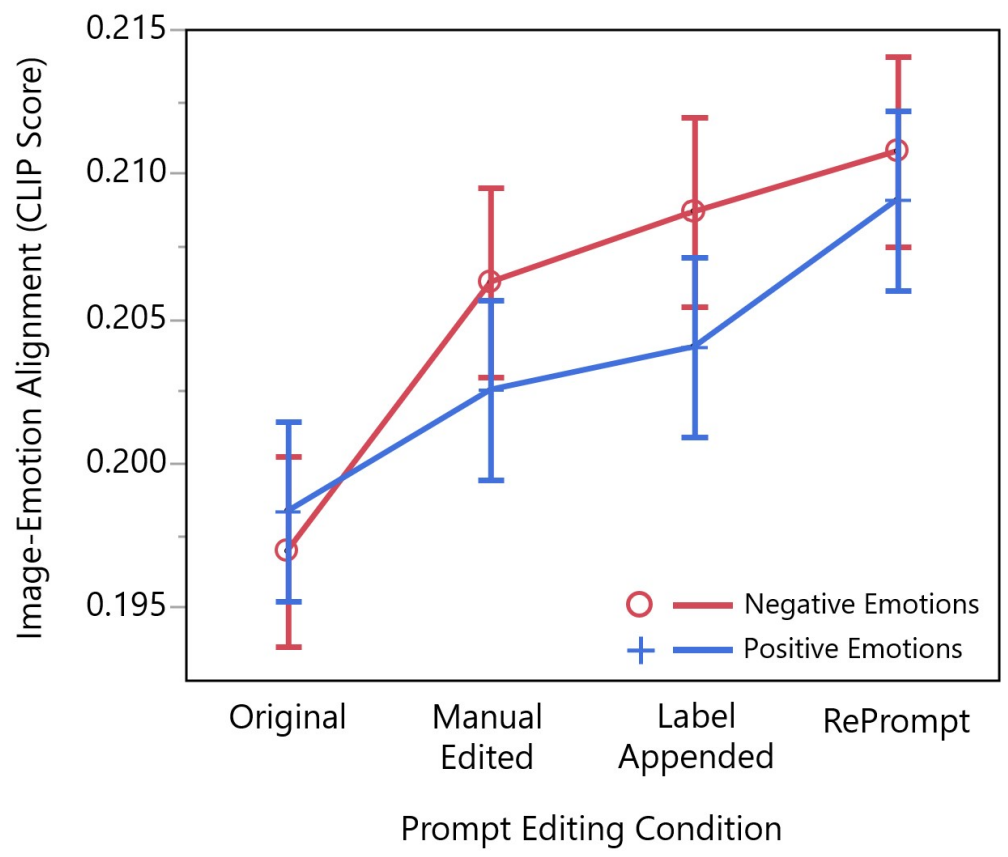

**Figure 9: Results of CLIP Score of Image-Emotion alignment with modeling Emotion Type and its interaction with Prompt Editing Condition as the fixed effect.**

Both IEA and ITA ratings for RePrompt images were significantly higher than other conditions for only negative emotions.

**Interpretation of IEA rating results.** There are three potential explanations for the results on IEA ratings. Firstly, participant raters might be more sensitive to images with negative emotions and less sensitive to differences in images with positive emotions. This explanation is supported by prior studies suggesting that humans are sensitive to valence differences in emotionally negative stimuli but not for positive stimuli [89]. According to human evolution theories, evolutionary pressure has led the nervous system to guarantee rapid and intense responses to negative events [11] and human brains have evolved routes for image processing as a survival mechanism [2]. Secondly, RePrompt did not improve the image quality of positive emotions sufficiently to be perceived by human raters. We fit another LMER model for the simulation study results by adding Emotion Type and its interaction with Prompt Editing Condition as a fixed main effect, but the statistical results did not show significant differences between positive emotion and negative emotion for both Original Prompt and RePrompt, shown in Figure 9 (see Table 5 in the Appendix for results of the ANOVA tests). Therefore, the second explanation is less likely to be true. Thirdly, CLIP Score might not model human perception of positive emotions well. To examine this hypothesis, we conducted a follow-up study with the Emotion6 [58] dataset that labeled each image with a probability distribution of six basic emotions (i.e., "sadness," "joy," "fear," "disgust," "anger," and "surprise"). We first computed CLIP Scores for each image for each of the emotion label, then calculated the Pearson $r$ correlation between CLIP Score and each emotion. Our results suggested that CLIP Score was more strongly correlated with negative emotions (r = 0.242, 0.348, 0.423, and 0.612 for "anger", "disgust", "fear", and "sadness", respectively) than positive ones (r = 0.076 and -0.159 for "joy" and "surprise", respectively). Table 8 in the Appendix includes the statistical details. In summary, the differences between our simulation and user study results might be related to peoples' lower sensitivity to positive stimuli and the weaker ability of CLIP in modeling positive emotions.

**Interpretation of ITA rating results.** The result of ITA ratings was inconsistent with our simulation study result. There are two possible explanations for the findings. Firstly, the edited texts in the RePrompt and Manual conditions resulted in images that kept the key meanings of the original texts; thus, raters did not perceive a lack of alignment between the images and the original texts. This suggests that the CLIP Score of ITA did not reflect human perception in our case. Secondly, raters' ITA ratings were influenced by their IEA ratings, as indicated by their strong correlation (Pearson r = 0.764, p < .0001). In our survey, ITA rating questions came after IEA, suggesting an order effect. Moreover, we checked the rating durations and found a significant difference (p < .0001) between IEA ratings (M = 35.7 s) and ITA ratings (M = 27.4 s), suggesting that raters spent less effort on ITA ratings possibly because they might follow their IEA ratings for ITA ratings. Both explanations were reasonable, so we concluded that the results of ITA were mixed and deserved further investigation in future work.

*6.4.4 Results of Rankings.* We did not find any significant difference across conditions for both IEA and ITA rankings, regardless of considering emotion type in the model. We found a statistically significant yet weak correlation between rating scores (higher is better) and ranking orders (lower is better) for both IEA (r = -0.114, p < .0001) and ITA (r = -0.110, p < .0001). Even though the ranking orders were not sensitive as rating scores, the ranking questions might have influenced participants' rankings.

## 7 SUMMARY OF RESULTS

**RQ1. How would laypersons perceive the AI-generated images regarding emotional expressiveness and strategically refine the text prompts to improve image generation?** Our interview participants found that the AI 1) could understand simple and concrete texts instead of complex and abstract texts; 2) did not understand some concepts; 3) understood objective descriptions better; and 4) was not quite accurate and sometimes difficult to understand when it fails. Participants' editing strategies primarily included: 1) emphasizing emotion by adding or adjusting emotion words or context; 2) simplifying text by deleting sentences or paraphrasing; 3) increasing text concreteness; and 4) changing self-expression from a first-person perspective to an image description from a third-person perspective.

**RQ2. How can text prompts be automatically refined to generate better images regarding emotional expressiveness?** In RePrompt, we curated intuitive text features based on layperson editing strategies, trained machine learning models to predict image quality scores with the text features, then applied model explanations to the trained model to generate a rubric for automatically editing the text prompts.

**RQ3. How effective is RePrompt at improving emotional expression in image generation?** The simulation study and user study results suggest that RePrompt improves image generation with the AI model with respect to IEA, especially for negative emotions. The validators in our evaluation user study could not perceive differences in the expression of positive emotions across conditions, which might be due to a lower sensitivity to positive stimuli in humans and CLIP's weaker ability in modeling positive emotions. The results were mixed for ITA.

# 8 DISCUSSION

## 8.1 Design Implications and Limitations

**For emotional text-to-image generation, the CLIP Score of image-text alignment (ITA) does not match human perception, while CLIP Score of image-emotion alignment (IEA) matches for only negative emotions.** Our results suggest that, for negative emotions, people could perceive different levels of IEA that were consistent with the CLIP Score. One reason could be the difference of human perception sensitivity between positive and negative images evidenced by neuroscience studies of human brain response to stimuli of different levels of emotionally positive and negative images [89]. We found a similar result in the AffectGAN study [22], where the images with negative emotions (i.e., anger and depression) were more accurately labeled by human validators than those with positive emotions (i.e., happiness and calmness). Another reason could be that CLIP could not well model nuanced positive emotions, as our validation study with another emotional image dataset suggested (Section 6.4.3). Future work should consider this effect when using the AI-generated images for applications like emotion regulation [28, 30, 78].

**Emotion bias in CLIP and text-to-image generative models**. Recent works have studied the bias (e.g., gender and skin tone) and ethical risks (e.g., harassment and violence) of using text-to-image generative models [15, 56]. Our findings suggest another bias—emotion valence sensitivity—in CLIP-based text-to-image generative models. This might be due to the lack of emotion datasets for training generative models or the lack of well-designed training strategies for modeling the nuances in positive emotions. Due to the complexity of emotions embedded in text and images (e.g., multiple emotions expressed in one image and emotion change with different contexts [58]), sophisticated datasets and modeling are warranted to reduce the bias and improve model sensitivity.

**Text-to-Image models may share transferable knowledge.** Our rubric was developed based on model explanations of the proxy model trained on the images generated by VQGAN-CLIP [19] for scalability and applied the edited prompts to DALL·E 2 [61] to generate images for user evaluation. Our results suggest that the knowledge we derived with VQGAN-CLIP could be transferred to image generation with DALL·E 2. We assume that the transferability should be due to CLIP [60] in both models. Future work could investigate the transferability among more models.

**Explainable AI could benefit prompt engineering.** The information loss in communication between human requestors and AI models is the fundamental reason behind the need for such prompt engineering to "bridge" or "translate" users' requests to generative AI models. Even for human-to-human communications, we may misunderstand each other during conversations due to inadequate emotional intelligence and empathy. With RePrompt, we proposed a novel explainable AI (XAI) approach for prompt engineering, which could improve human-AI communication by tuning human requests to better prompt the AI model to generate more precise output. The model explanation on the proxy model is a new method of applying XAI.

**Limitations on feature selection, modeling, and emotional expression measurement.** Although we developed an intuitive and effective rubric for automatic prompt-editing, our feature selection and modeling methods were simplistic. Firstly, our features were derived based on individual words in the text, which might break phrases and meaningful chunks. For example, we observed that RePrompt would break "old friends" into "old, friends" so that DALL·E 2 generated images portraying many elderly people, even though "old friends" might not necessarily mean elderly friends. Adding more sophisticated, but difficult-to-understand, features like n-grams, affective lexicon, and dependency parsing, might increase the performance of our approach, which we defer to future exploration. Secondly, our modeling considered only text semantics without image properties like styles and color themes. In the use case of news illustration with text-to-image models, Opal [43] suggested art styles for users to choose according to the news' contents and tones. Future work of RePrompt should mitigate these limitations. Thirdly, even though we adopted subjective ratings with crowd workers to measure the human perception of emotional expression in images, we note that there are multiple measures to evaluate human's emotional reaction to visual stimuli. For more rigorous investigation, measures with electrocardiogram [76], facial electrocardiogram [23], and skin conductance variability [77] should be utilized.

## 8.2 Generalization of RePrompt

We discuss the generalization of our method from two perspective: 1) the generalization of the rubric generated in our study (see Section 5.5) to other emotional datasets, and 2) the generalization of the RePrompt pipeline to other types of expressions in text-to-image generative models as well as other generative models.

**Generalization of the rubric for emotional image generation**. As our study results suggested, the rubric we derived with VQGAN-CLIP was still effective with DALL·E 2 for improving emotional expressiveness. However, we used the same EmpatheticDialogues dataset for both evaluations. It would be worthwhile to examine if the rubric is limited to the specific dataset, so we tested on another emotional text dataset curated from Twitter [68]. We applied the same method with our rubric as described in Section 5.5 to revise the original texts, then generated the images with original texts, label appended texts, and RePrompt edited texts, respectively. The new results of the CLIP Score for IEA align with our prior results that RePrompt edited prompts could outperform other conditions, suggesting that the rubric is generalizable to other datasets. See details of this experiment in Appendix A.5.

**Generalization of the RePrompt pipeline**. The three key components in RePrompt are feature curation, proxy modeling, and model explanations. The proxy model is a lightweight proxy to the generative model that requires explanations. Applying model explanations to the proxy model enables feature analysis that could be transferred to the target generative AI model. In our use case, the explanations guide the development of automatic prompt-editing for better output by the generative model. When adapted to other domains, the proxy model (e.g., LightGBM [34]), feature set (e.g., number and concreteness of nouns), and model explanations could be adjusted with many options. For example, besides the PDP, other counterfactual XAI techniques such as Anchors [64] and Scalable Bayesian rule lists [86] are available. To enable human-readable prompt engineering, selecting intuitive and effective features is crucial and a user-centered method as in our interview study might be necessary. In theory, the RePrompt pipeline could be generalized to other generative models besides text-to-image models and other types of expressions. For example, if a music generative model takes lyrics or videos as prompts, one might use RePrompt to investigate which features of the lyrics (e.g., word concreteness) or video (e.g., colors, objects) could affect the output music and then develop a rubric for prompt engineering.

### 8.3 Benefits, Potential Applications, and Future Work

Our method provides human-readable prompt revisions, which can help users understand the model and reduce trial-and-error attempts. In a deployed setting, we could show users explicit edits (added or deleted words) to help their ideation for iterative prompt revision. We defer the further study of iteratively prompting with human-AI collaboration to future work. Additionally, our use case of emotional expression in images encourages future work to study potential applications for mental health. Emotion expression is a basic human need and critical for human health and wellbeing [69]. One could convey emotions in many ways (e.g., verbal language, body language, facial expressions, images, and music), among which visual art is a powerful and implacable one to express and perceive emotions. Text-to-image generative models have dramatically lowered the threshold for laypersons to create visual art for expressing emotions, while the proposed RePrompt method could improve the emotional expressiveness of the generative models, especially nuances in negative emotions. Therefore, RePrompt could potentially be used to support emotional expression via image creation. Prior studies have shown that appropriate photos and visual art could support brain health [5, 48] and aid psychological counseling by assisting the client in reflection and empowerment to change [24]. Therefore, it is worth investigating how RePrompt could support laypersons in generating personalized emotional images (e.g., using text description of personal experiences) for emotion communication and regulation (e.g., for stress-reduction [30]). Another application could be to enhance expressive writing by automatically generating contextual images. Prior research has suggested that expressive writing (i.e., writing about traumatic, stressful, or emotional events) could improve both mental and physical health for people with an earlier traumatic experience, pain and physical health in cancer, and poor sleepers, etc. [3]. Providing contextual images during the writing process could increase writers' physiological arousal and reflection, which could in turn support self-expression and ideation in writing. Nevertheless, we recall the limitation on emotional expression with the used models due to the emotion bias in CLIP and call for further study on user experience of using such tools for emotional expression towards realizing the potential applications.

## 9 CONCLUSION

We have proposed RePrompt to automatically refine text prompts for text-to-image generative models to improve the semantic precision of image generation. We first conducted an interview study to understand how laypersons perceived the AI-generated images and how they strategically edit the text prompts. Inspired by participants' prompt-editing strategies, we developed the RePrompt pipeline and created a rubric to allow explainable and automatic prompt-editing to improve the emotional expressiveness of image generation in generative AI models. Through a simulation study with objective measures and a user study with subjective ratings, our results suggested that the text prompts edited by RePrompt could nudge the generative AI models to generate better images in terms of image-emotion alignment, especially for negative emotions.

## ACKNOWLEDGMENTS

This work was supported by the Singapore Ministry of Education (MOE) Academic Research Fund Tier 2 (T2EP20121-004) and NUS iHealthtech Smart Sensors and Artificial Intelligence (AI) for Health grant.

# A APPENDIX

## A.1 A.1 Examples of CLIP Scores

We list CLIP Score examples in Table 3 below. We note that CLIP could generate embeddings for any textual and image input, including junk input (e.g., "#%@)%I"), leading to meaningless CLIP Scores, which should be avoided when applying the CLIP Score. For example, the CLIP Score of "#%@)%I" and "dog" is 0.8869, which is higher than the CLIP Score (0. 8832) of "animal" and "dog". The CLIP Score of "!@#$%" and the left image in the first row in Table 3 are 0.1826, which is even higher than the IEA Score. The problem of junk characters can be addressed by checking word spelling. We checked the used text dataset in our study and found no such junk input.

**Table 3: Example CLIP Scores for image-emotion alignment (IEA) and image-text alignment (ITA) of images generated by DALL·E 2.**

| Emotion & Text | Image and CLIP Scores | | |
|---|---|---|---|
| Emotion: joyful<br>Text: I'm pretty happy that my daughters have a day off from school today. | 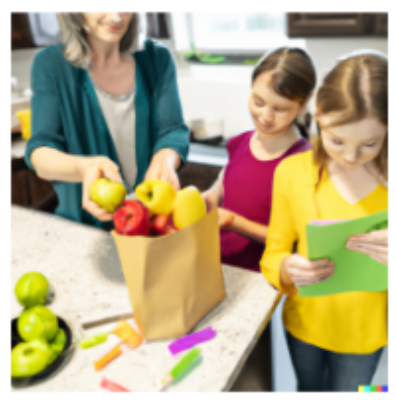<br>IEA = 0.1795<br>ITA = 0.2068 | 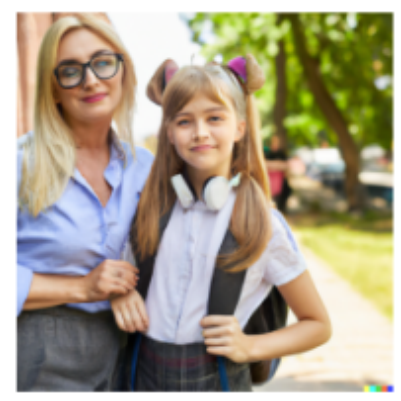<br>IEA = 0.1895<br>ITA = 0.2354 | 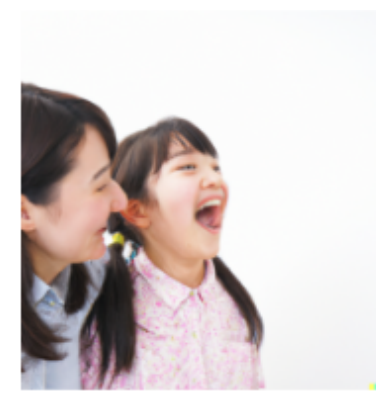<br>IEA = 0.2455<br>ITA = 0.2433 |
| Emotion: proud<br>Text: My son got straight As in his summer class and it makes me happy. | 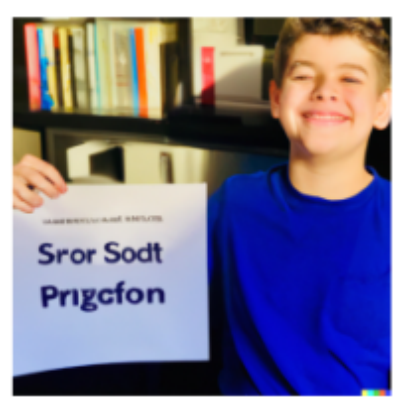<br>IEA = 0.2163<br>ITA = 0.2653 | 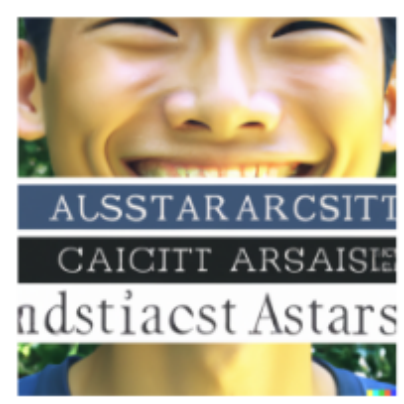<br>IEA = 0.1957<br>ITA = 0.2287 | 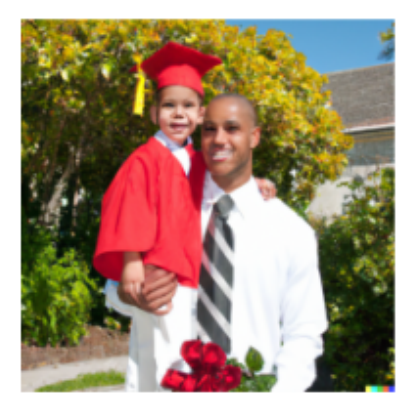<br>IEA = 0.2238<br>ITA = 0.2457 |
| Emotion: lonely<br>Text: Went to the skating rink alone yesterday. Wasn't too grand. | 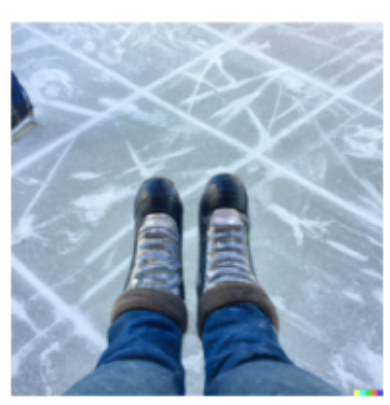<br>IEA = 0.2070<br>ITA = 0.2955 | 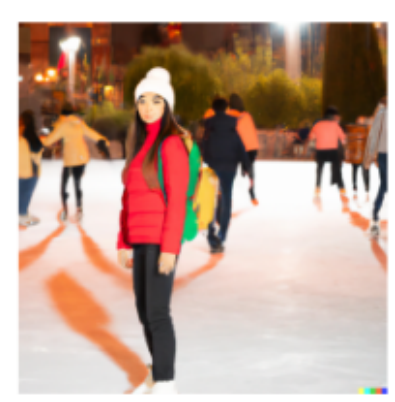<br>IEA = 0.2164<br>ITA = 0.2747 | 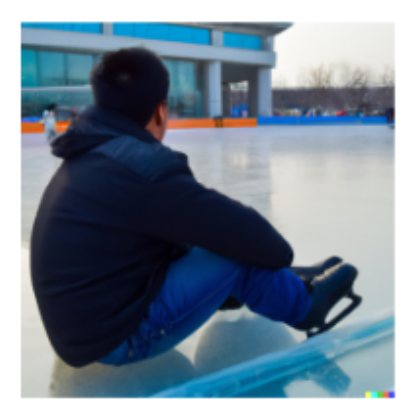<br>IEA = 0.2468<br>ITA = 0.3359 |
| Emotion: angry<br>Text: I'm pretty mad about the fact my internet bill went up 40 dollars out of nowhere | 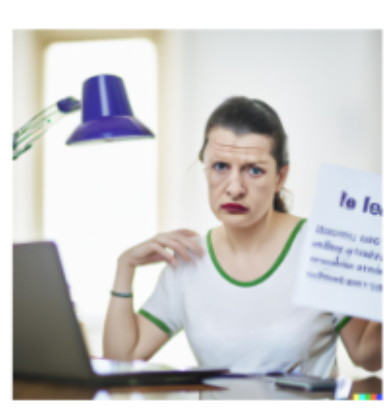<br>IEA = 0.2060<br>ITA = 0.1775 | 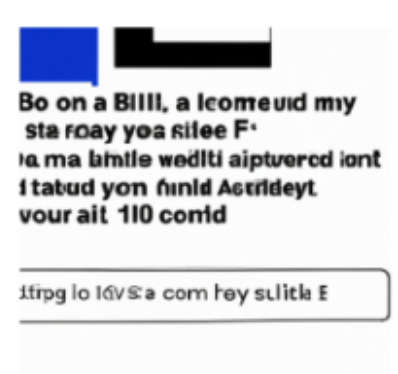<br>IEA = 0.2195<br>ITA = 0.2473 | 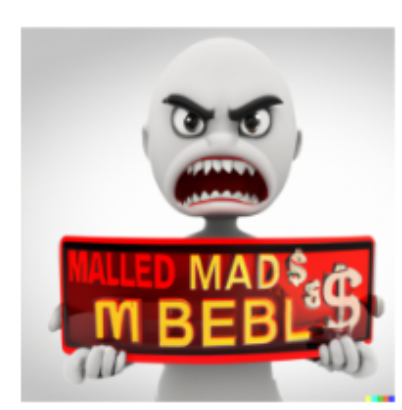<br>IEA = 0.2661<br>ITA = 0.1779 |

## A.2 Model Explanations

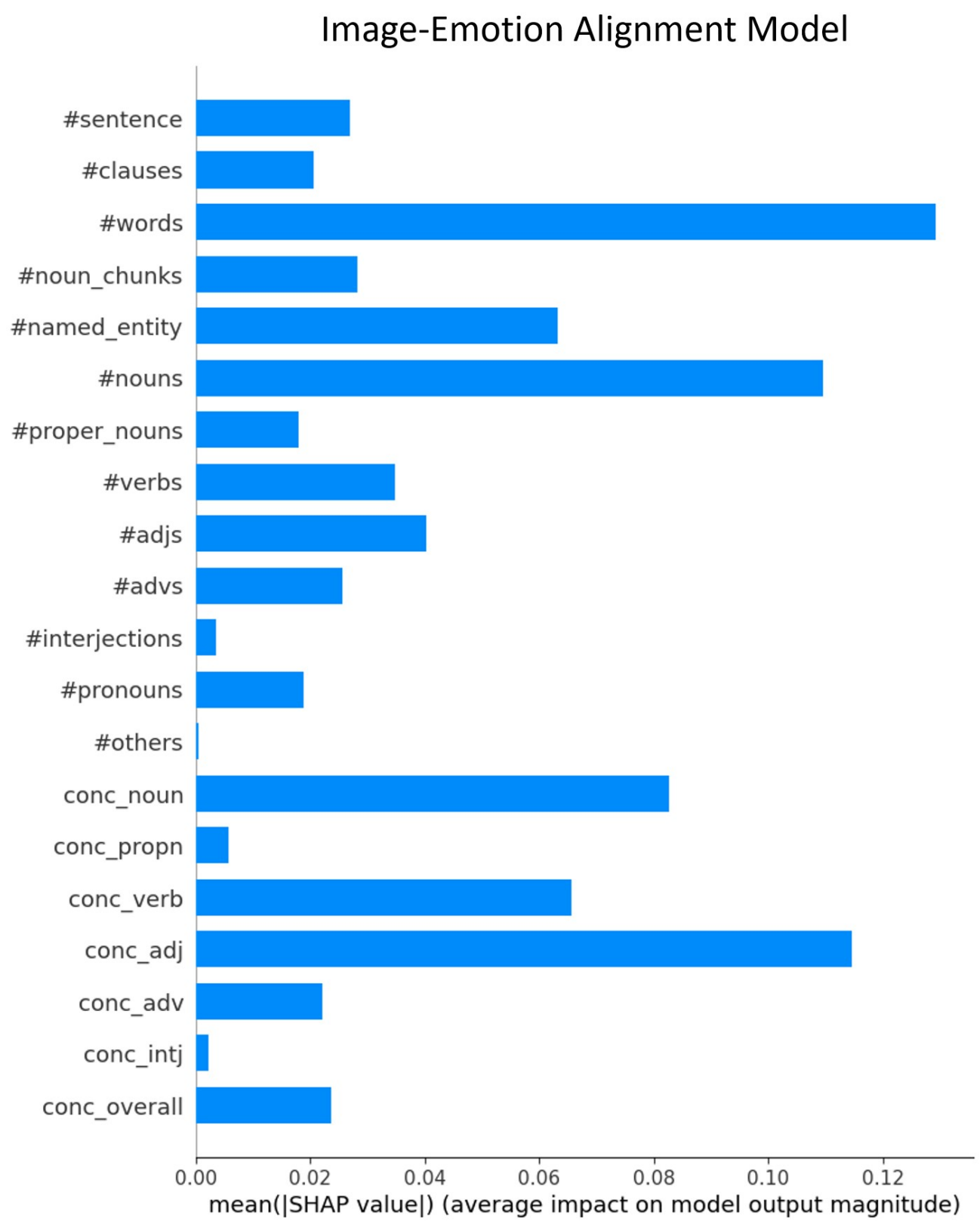

**Figure 10: Feature importance according to SHAP values in the image-emotion alignment model.**

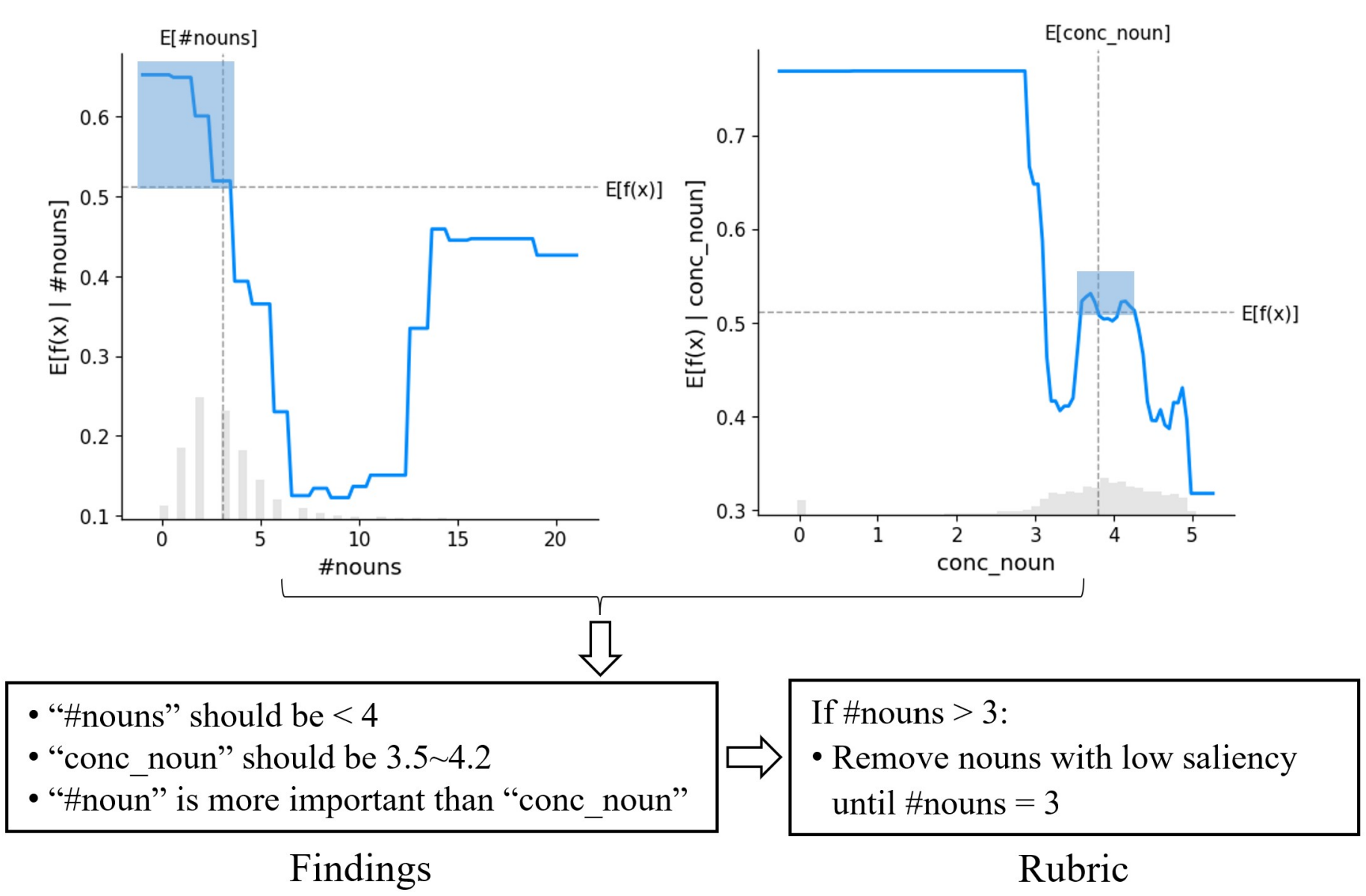

**Figure 11: Rubric curation for nouns.**

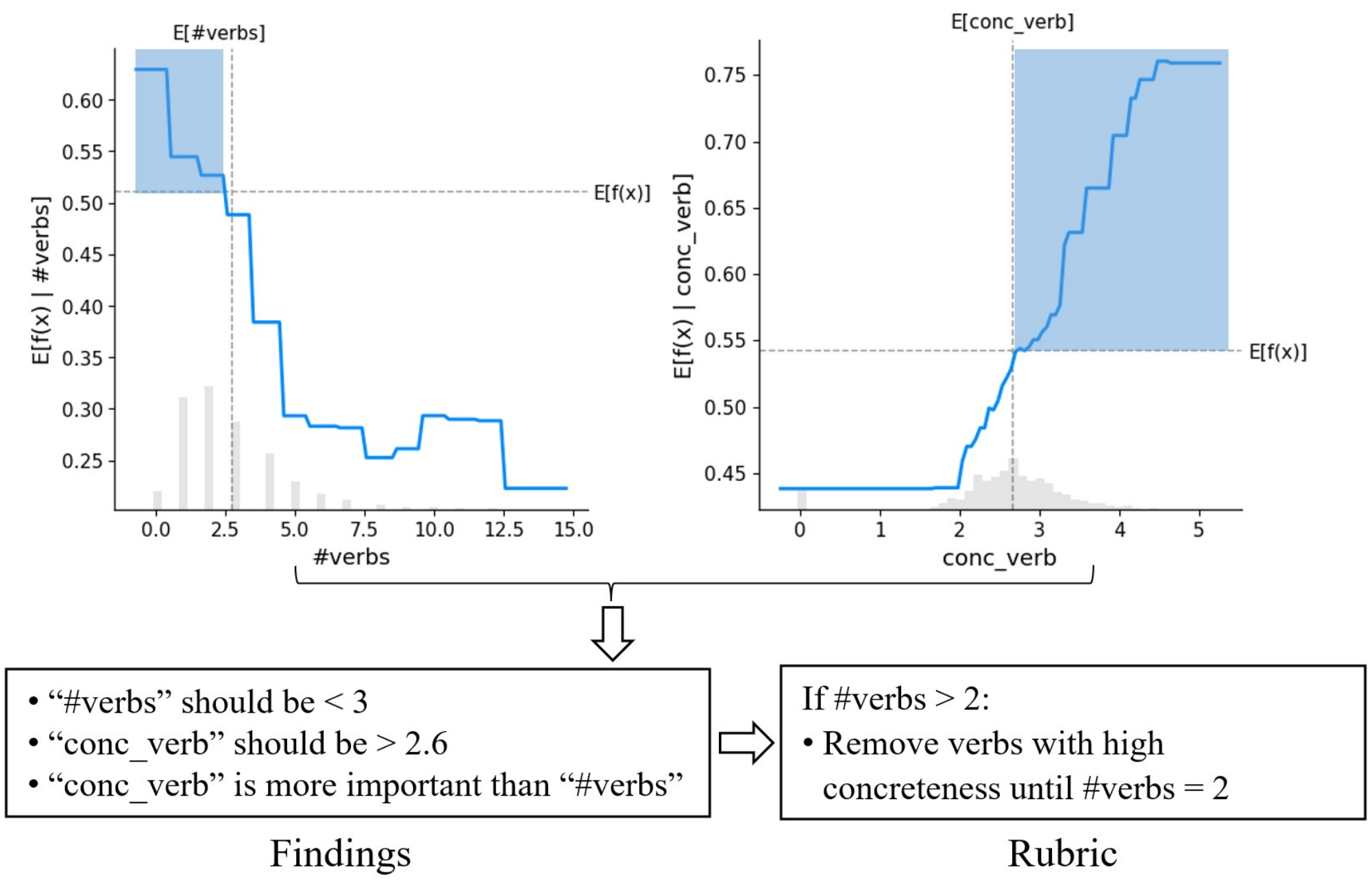

**Figure 12: Rubric curation for verbs.**

## A.3 Survey Screenshots in the Image Evaluation User Study

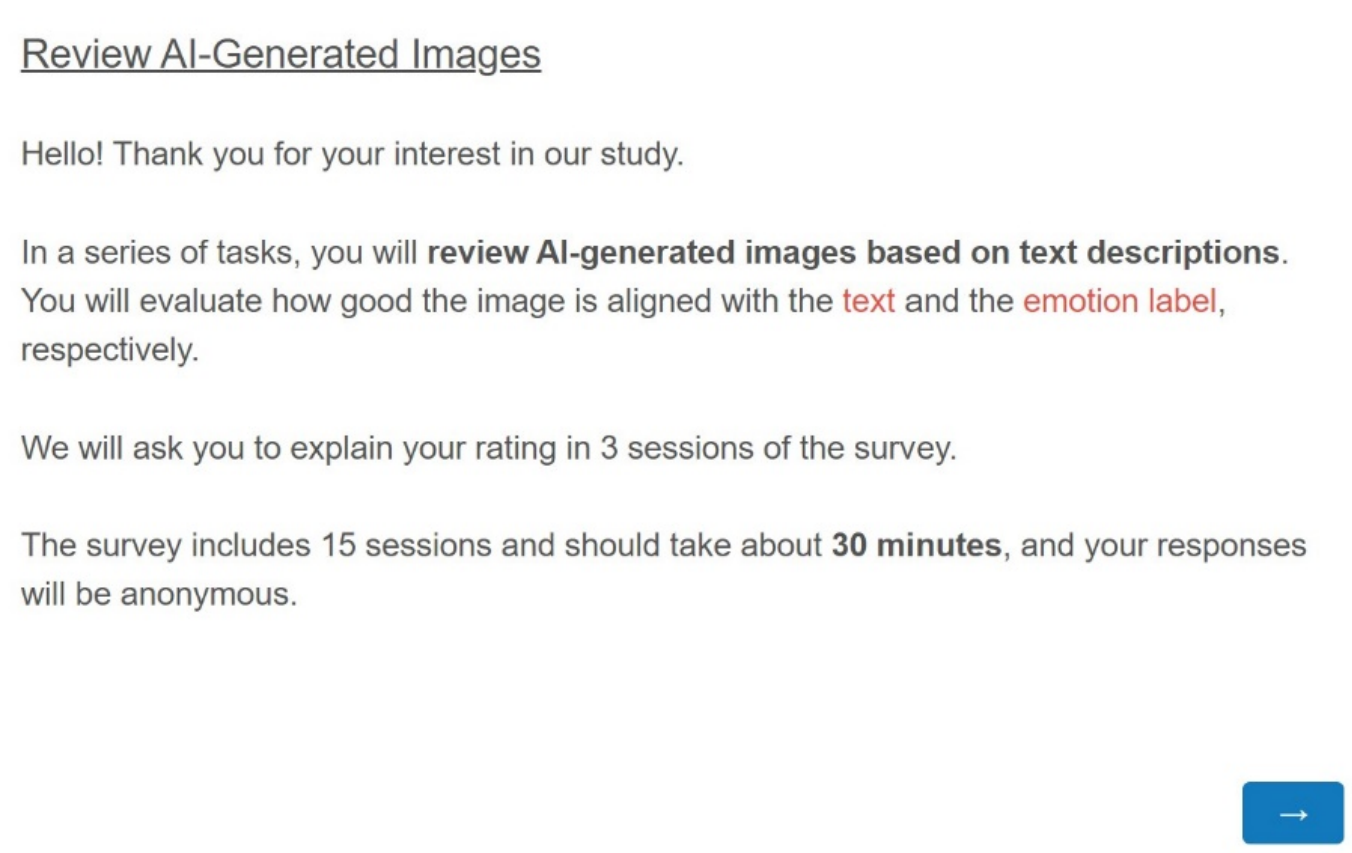

**Figure 13: Survey introduction.**

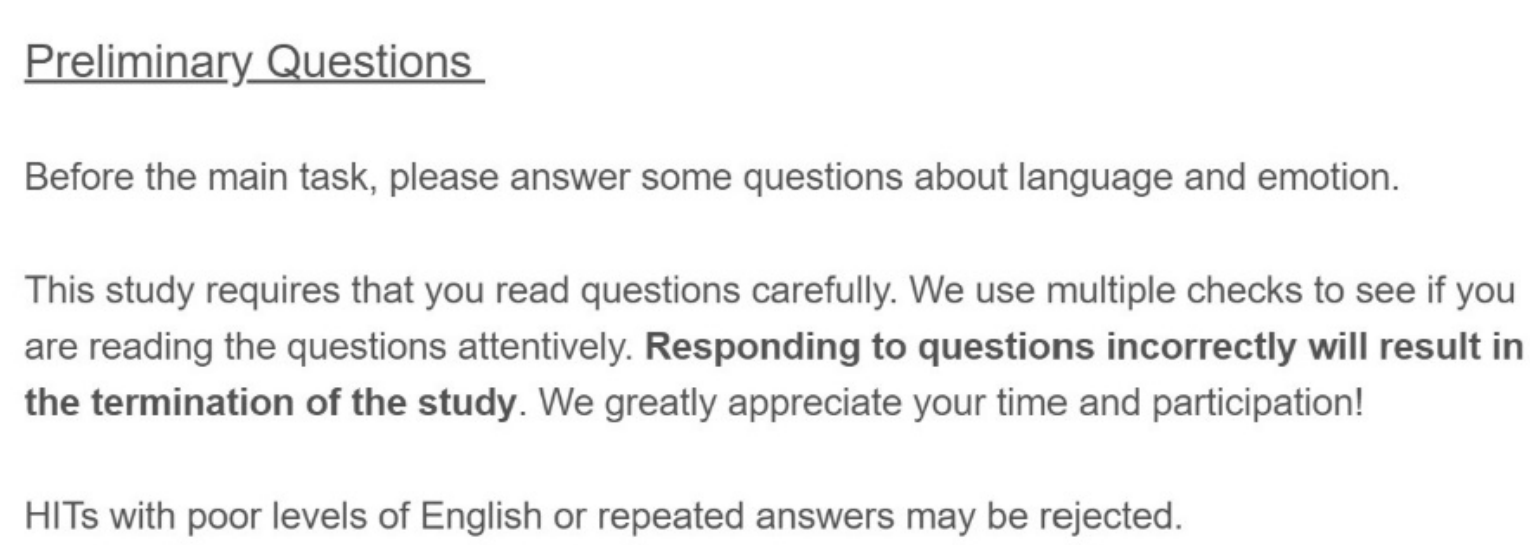

**Figure 14: Screening questions introduction.**

**Language Questions:**

Which of the following words is MOST related to "moody"?

- distant
- stable
- fantastic
- emotional

Which of the following words is MOST related to "distracted"?

- thoughtful
- unfocused
- generous
- beautiful

Figure 15: Language questions (Part 1).

Which of the following words is MOST related to "persevere"?

- persist
- tame
- forgetful
- lucky

Which of the following words is MOST related to "sympathy"?

- compassion
- sociable
- truthful
- honest

→

Figure 16: Language questions (Part 2).

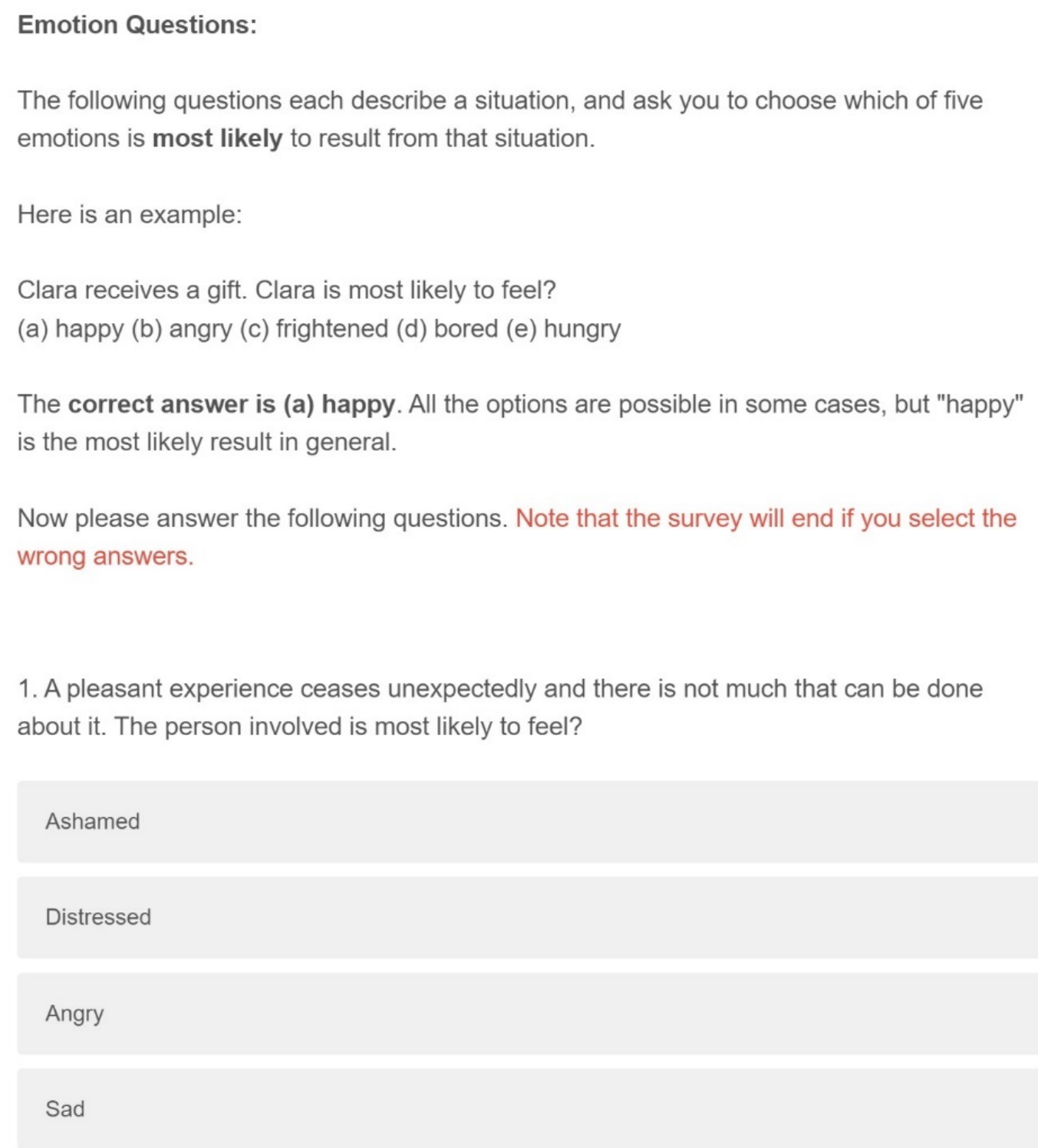

**Emotion Questions:**

The following questions each describe a situation, and ask you to choose which of five emotions is **most likely** to result from that situation.

Here is an example:

Clara receives a gift. Clara is most likely to feel?
(a) happy (b) angry (c) frightened (d) bored (e) hungry

The **correct answer is (a) happy**. All the options are possible in some cases, but "happy" is the most likely result in general.

Now please answer the following questions. Note that the survey will end if you select the wrong answers.

1. A pleasant experience ceases unexpectedly and there is not much that can be done about it. The person involved is most likely to feel?

Ashamed

Distressed

Angry

Sad

**Figure 17: Emotion questions (Part 1).**

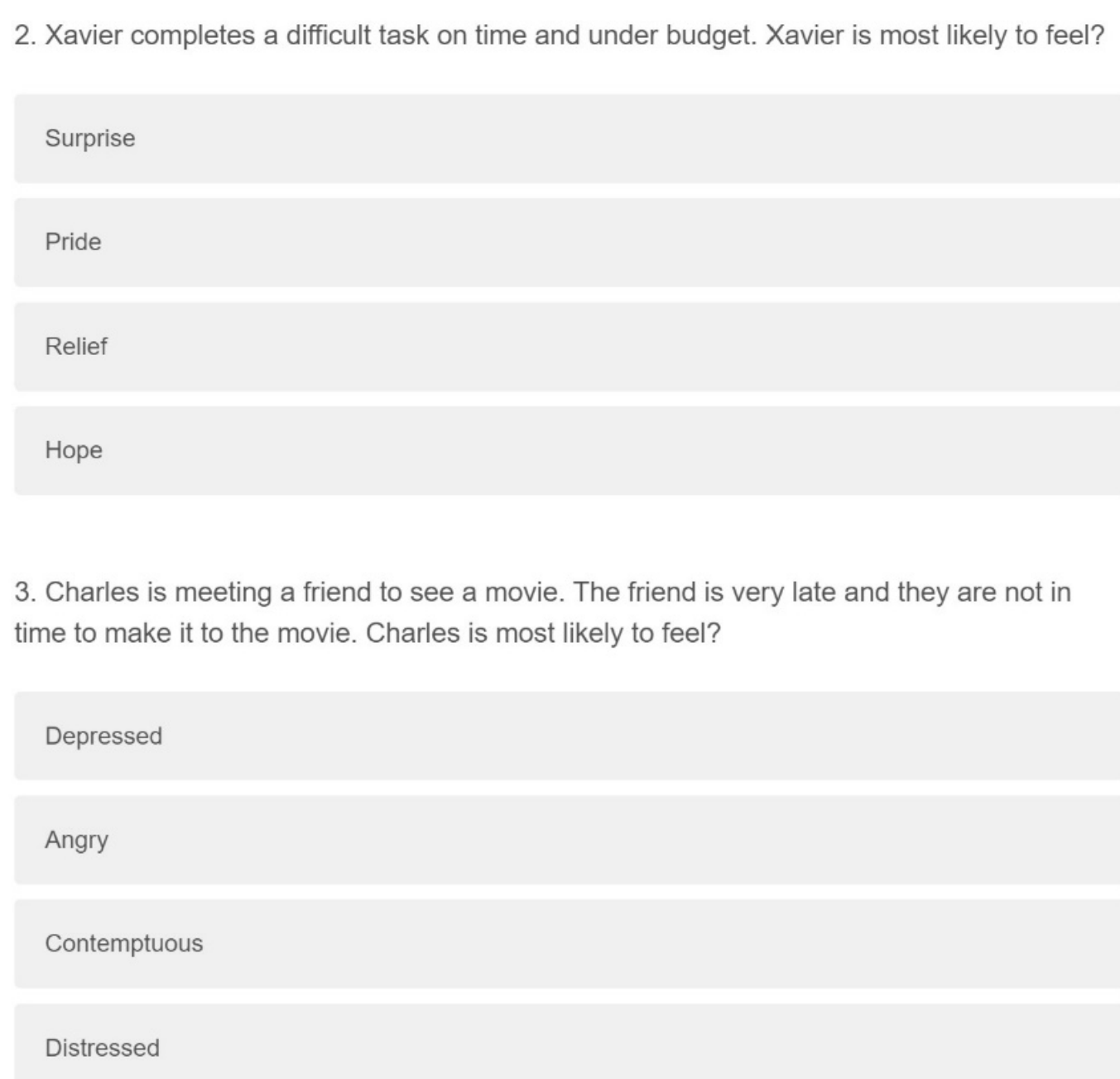

2. Xavier completes a difficult task on time and under budget. Xavier is most likely to feel?

Surprise

Pride

Relief

Hope

3. Charles is meeting a friend to see a movie. The friend is very late and they are not in time to make it to the movie. Charles is most likely to feel?

Depressed

Angry

Contemptuous

Distressed

**Figure 18: Emotion questions (Part 2).**

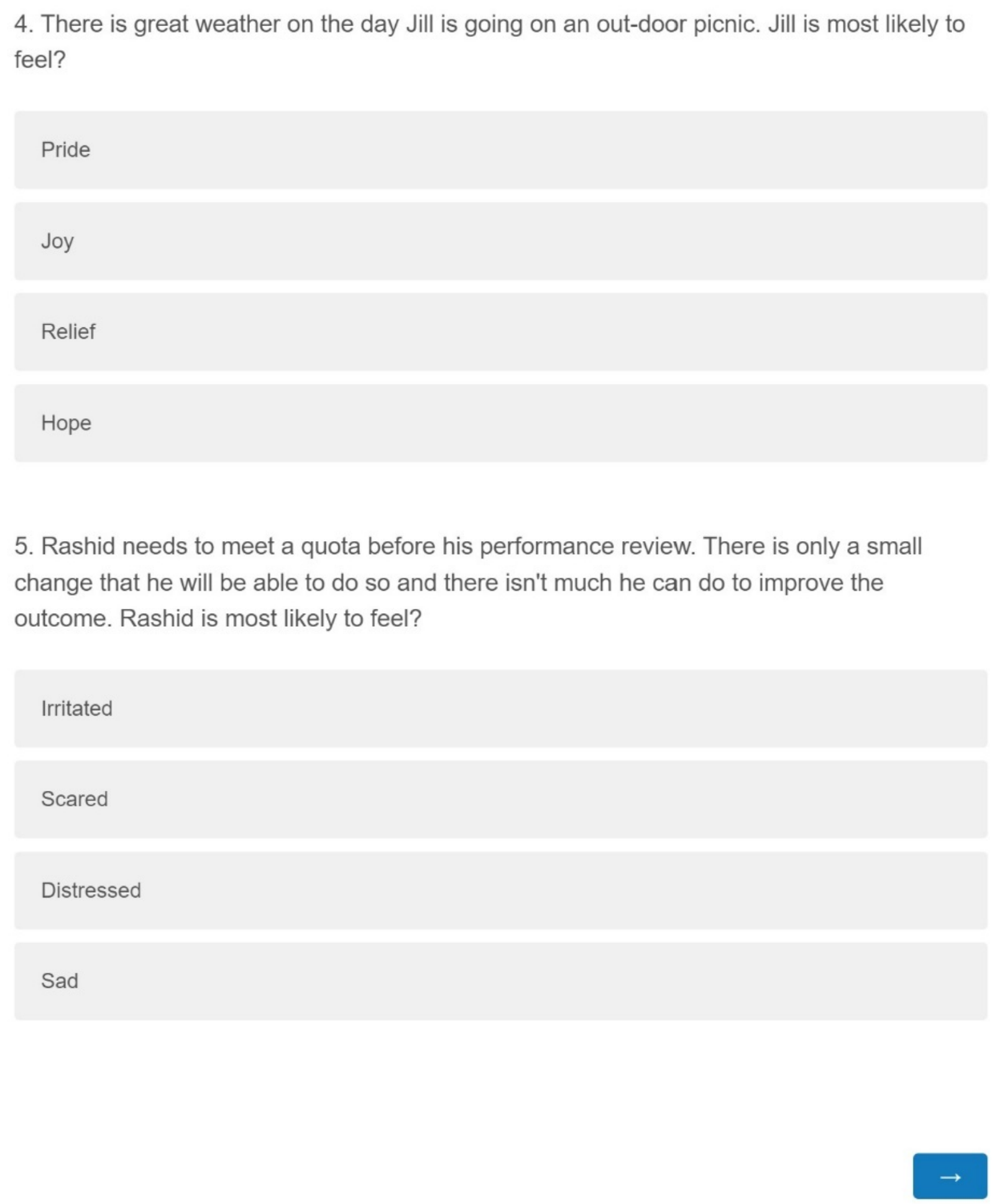
4. There is great weather on the day Jill is going on an out-door picnic. Jill is most likely to feel?

Pride

Joy

Relief

Hope

5. Rashid needs to meet a quota before his performance review. There is only a small change that he will be able to do so and there isn't much he can do to improve the outcome. Rashid is most likely to feel?

Irritated

Scared

Distressed

Sad

→

**Figure 19: Emotion questions (Part 3).**

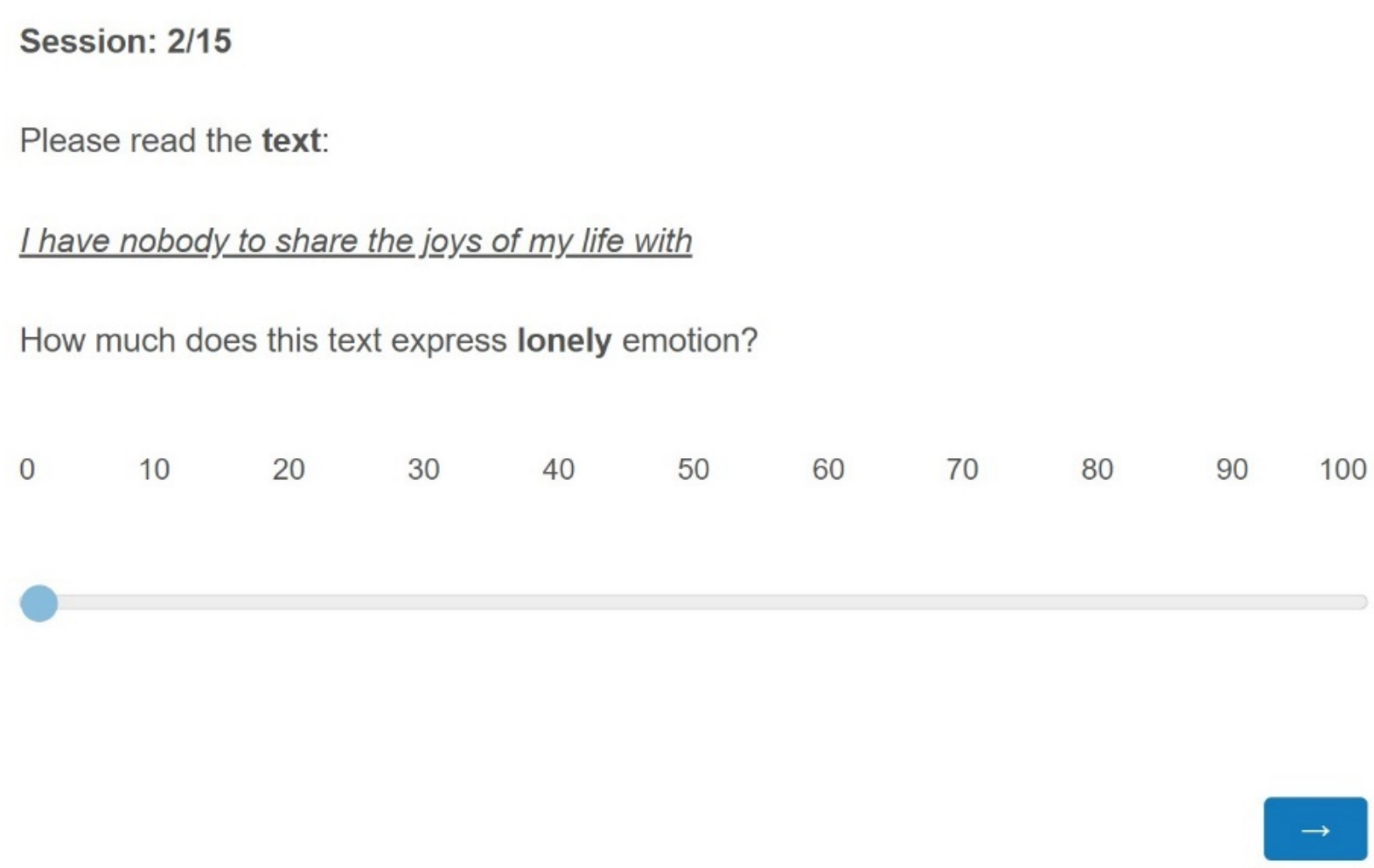
**Session: 2/15**

Please read the **text**:

*I have nobody to share the joys of my life with*

How much does this text express **lonely** emotion?

0 10 20 30 40 50 60 70 80 90 100

→

**Figure 20: Rating of text-emotion alignment.**

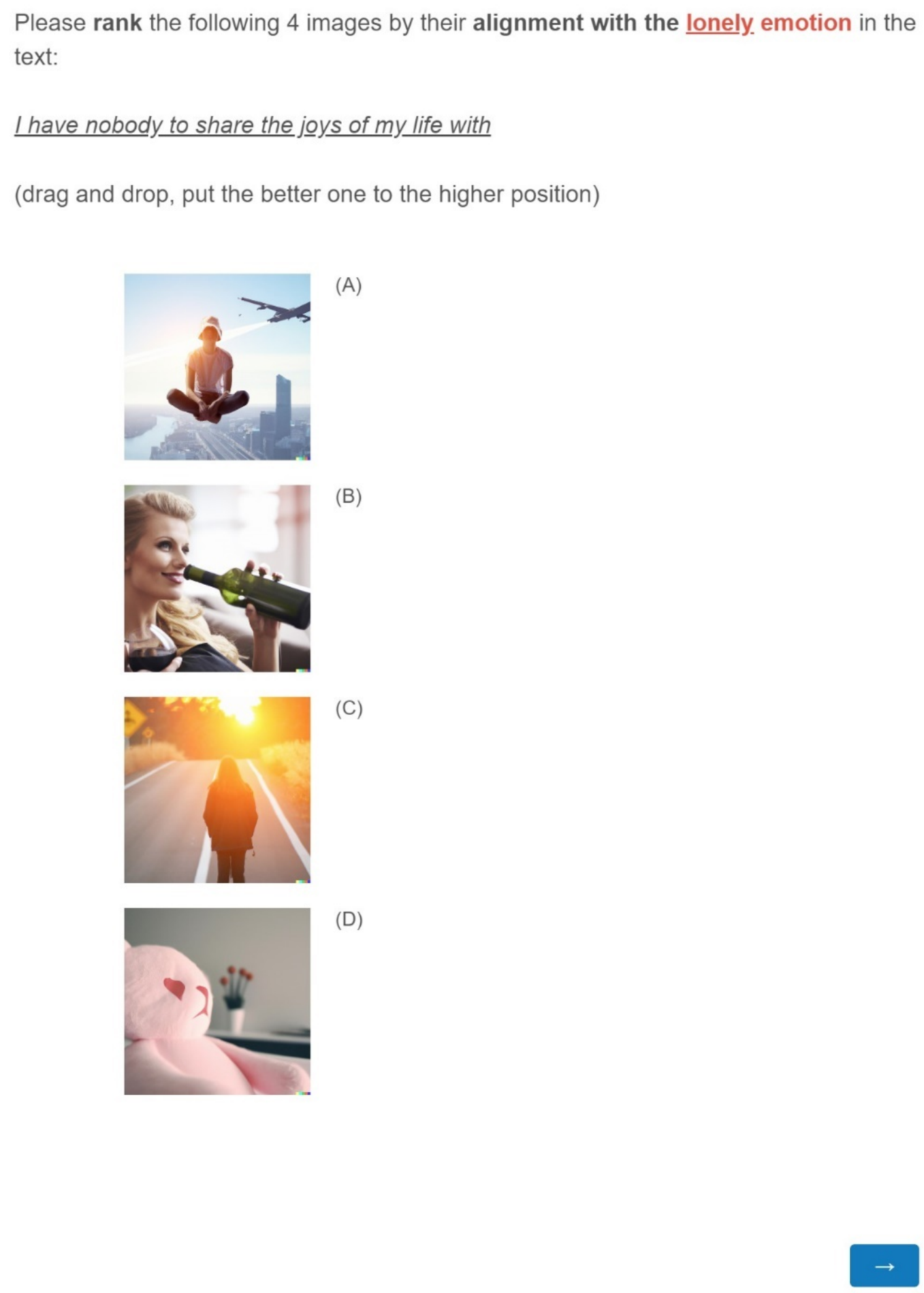

**Figure 21: Ranking of image-emotion alignment.**

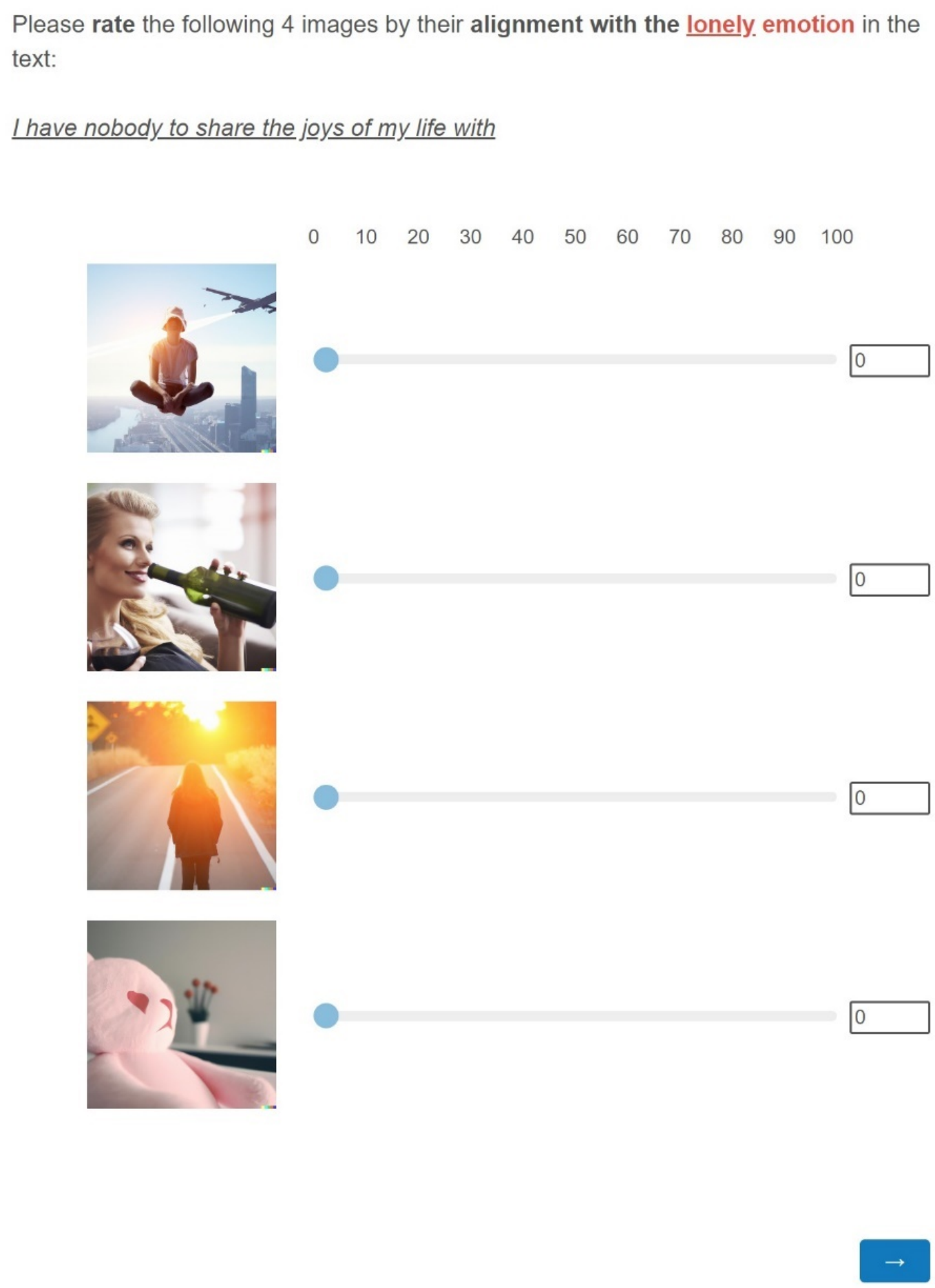

Figure 22: Rating of image-emotion alignment.

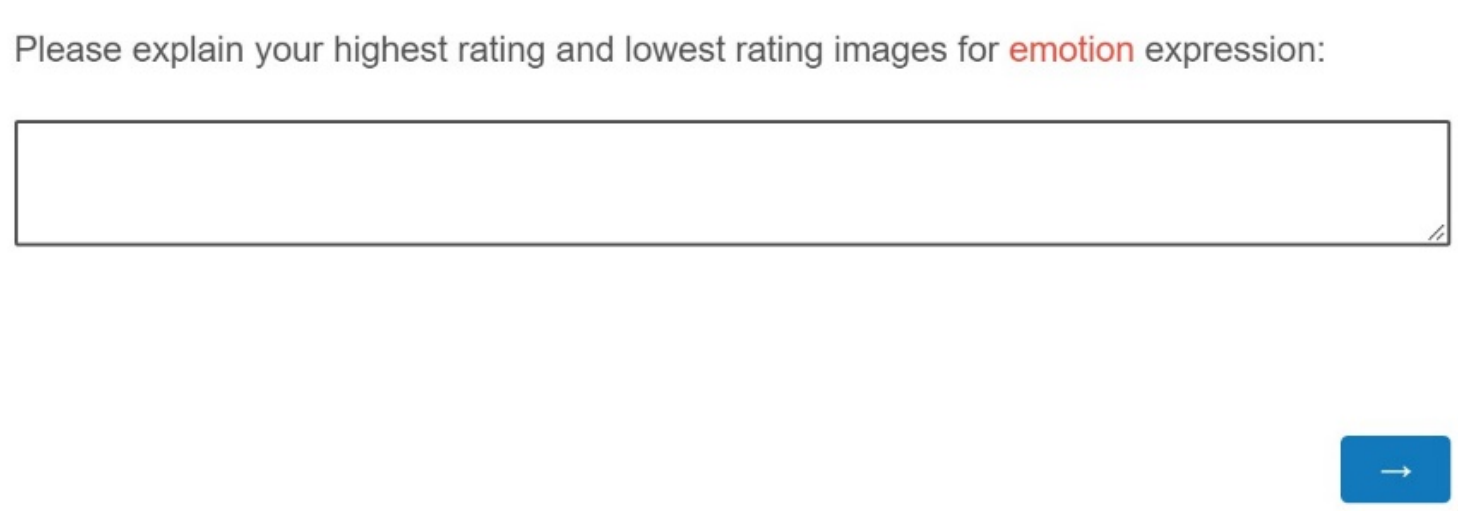

Figure 23: Explaining the ratings of image-emotion alignment.

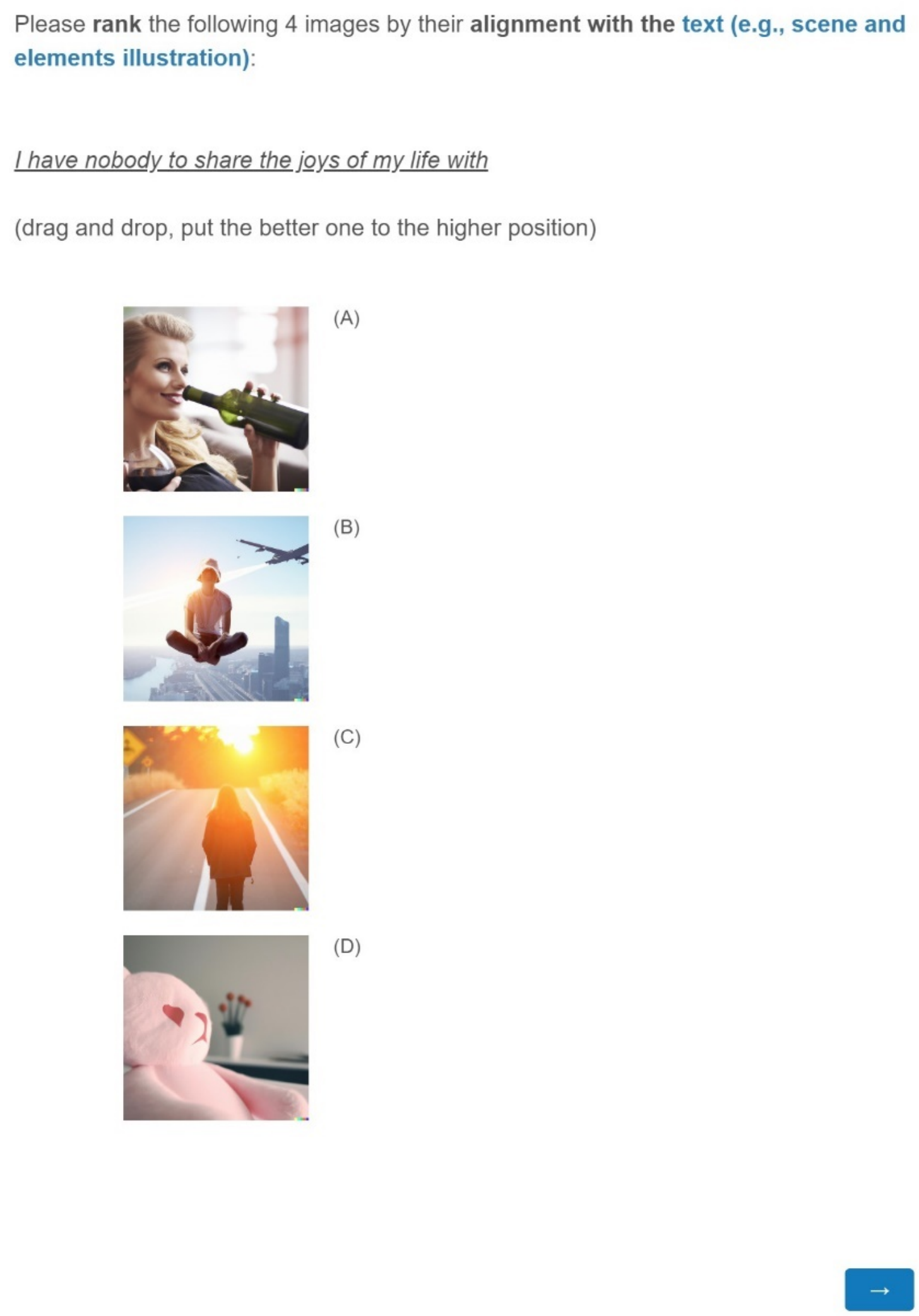

**Figure 24: Ranking of image-text alignment.**

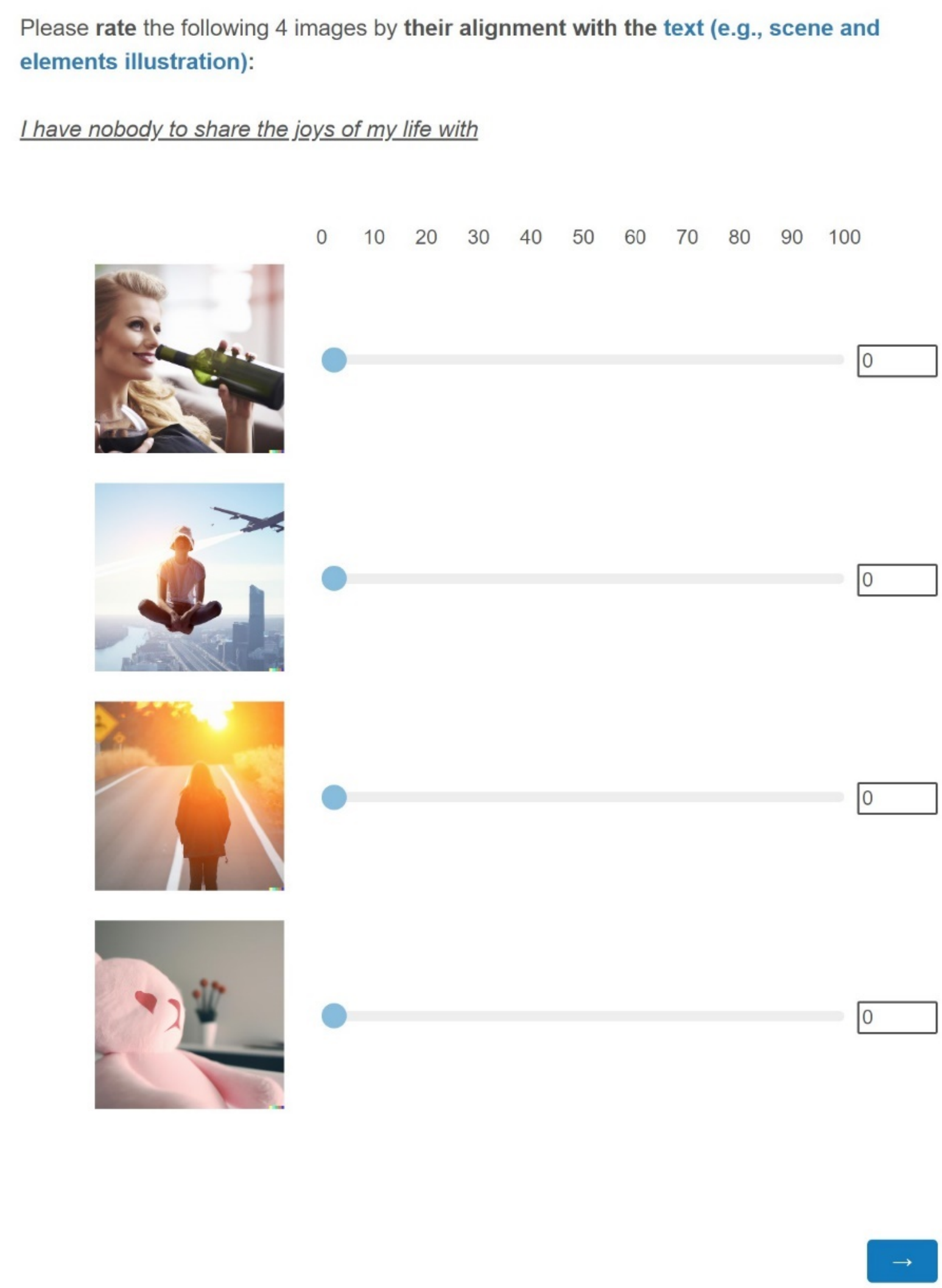

Figure 25: Rating of image-text alignment.

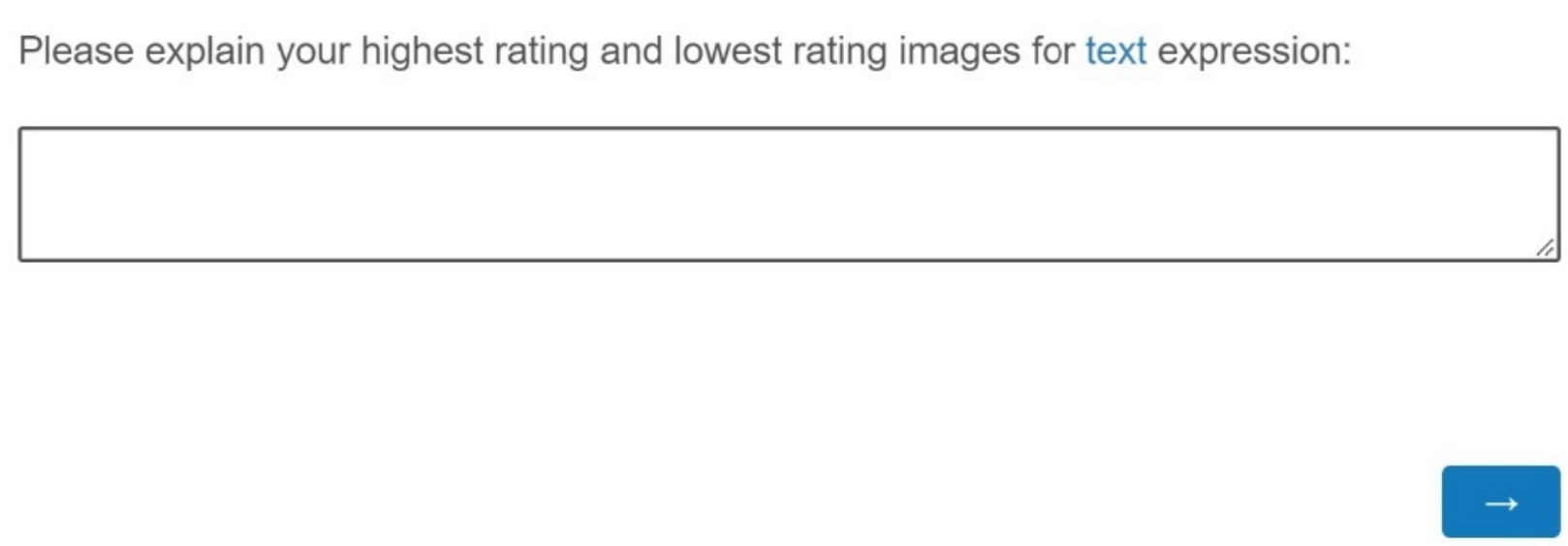

Figure 26: Explaining the rating of image-text alignment.

### A.4 Linear Mixed Models and Statistical Analysis Results

**Table 4: Statistical analysis for simulation study, with Prompt Editing Condition and Emotion as the main fixed effects and Prompt ID as the random effect. $p > F$ is the significance level of the fixed effect ANOVA. $R^2$ is the model's coefficient of determination to indicate goodness-of-fit.**

| Response | Linear Effects Model | $p > F$ | $R^2$ |
|---|---|---|---|
| Image-Emotion Alignment (CLIP Score) | Prompt Editing Condition | <.0001 | .35 |
| Image-Text Alignment (CLIP Score) | Prompt Editing Condition | <.0001 | .56 |

**Table 5: Statistical analysis for simulation study, with Prompt Editing Condition and Emotion as the main fixed effects and Prompt ID as the random effect. $p > F$ is the significance level of the fixed effect ANOVA. $R^2$ is the model's coefficient of determination to indicate goodness-of-fit.**

| Response | Linear Effects Model | $p > F$ | $R^2$ |
|---|---|---|---|
| Image-Emotion Alignment (CLIP Score) | Prompt Editing Condition | <.0001 | .35 |
| | Emotion Type | .0132 | |
| | Condition*Emotion Type | .0720 | |
| Image-Text Alignment (CLIP Score) | Prompt Editing Condition | <.0001 | .14 |
| | Emotion Type | <.0001 | |
| | Condition*Emotion Type | .0002 | |

**Table 6: Statistical analysis for user evaluation study, with Prompt Editing Condition and Emotion as the main fixed effects and Participant ID as the random effect. $p > F$ is the significance level of the fixed effect ANOVA. $R^2$ is the model's coefficient of determination to indicate goodness-of-fit.**

| Response | Linear Effects Model | $p > F$ | $R^2$ |
|---|---|---|---|
| Image-Emotion Alignment (Rating) | Prompt Editing Condition | <.0001 | .35 |
| Image-Text Alignment (Rating) | Prompt Editing Condition | <.0001 | .35 |

**Table 7: Statistical analysis for user evaluation study, with Prompt Editing Condition and Emotion as the main fixed effects and Participant ID as the random effect. $p > F$ is the significance level of the fixed effect ANOVA. $R^2$ is the model's coefficient of determination to indicate goodness-of-fit.**

| Response | Linear Effects Model | $p > F$ | $R^2$ |
|---|---|---|---|
| Image-Emotion Alignment (Rating) | Condition | <.0001 | .36 |
| | Emotion Type | <.0001 | |
| | Condition*Emotion Type | <.0001 | |
| Image-Text Alignment (Rating) | Condition | <.0001 | .36 |
| | Emotion Type | <.0001 | |
| | Condition*Emotion Type | <.0001 | |

**Table 8: Results of correlation between CLIP score and the probability distribution of emotion ratings from the Emotion6 dataset [58].**

| Emotion | Pearson' r | Count | Lower 95% | Upper 95% | $p$ |
|---|---|---|---|---|---|
| Anger | 0.2420 | 1980 | 0.2000 | 0.2830 | <.0001 |
| Fear | 0.4232 | 1980 | 0.3864 | 0.4587 | <.0001 |
| Sadness | 0.6117 | 1980 | 0.5834 | 0.6386 | <.0001 |
| Disgust | 0.3481 | 1980 | 0. 3088 | 0. 3862 | <.0001 |
| Joy | 0.0761 | 1980 | 0.0321 | 0.1197 | .0007 |
| Surprise | -0.1586 | 1980 | -0.2013 | -0.1154 | <.0001 |

## A.5 Statistical Analysis Results of Applying the Curated Rubric to Addition Dataset

We applied the rubric generated with RePrompt pipeline in our study (see Table 2 and Figure 5 in Section 5.5) to another emotional text dataset curated from Twitter [68]. This dataset contained six basic emotion types, namely sadness, joy, anger, love, surprise, and fear. We randomly selected 200 texts from each emotion type to make it a balanced test set. Then, we applied the same method in our main study to generate the RePrompt texts and Label Appended texts as the prompts, which resulted in corresponding images by the VQGAN-CLIP model with the same setting in our main study. The results of the image-emotion alignment are similar to the results based on the EmpatheticDialogues dataset, suggesting that the rubric is robust to apply to other datasets for emotional image generation. The statistical details are shown in Figure 27 and Table 9.

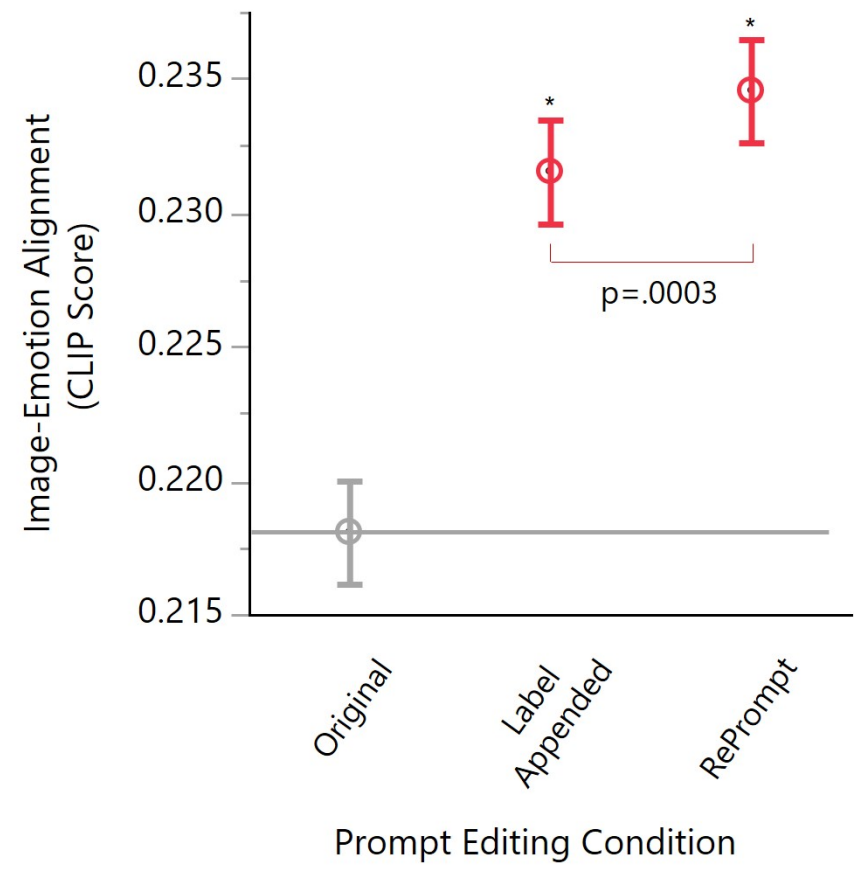

**Figure 27: Results of image-emotion alignment (IEA) on the generated images with the external emotional dataset [68]. P-values indicate difference from the Original condition if not specified. * indicates p < .0001. The goodness-of-fit of the LMER models and statistical significance of ANOVA tests are shown in Table 9.**

**Table 9: Statistical analysis for evaluation user study, with Prompt Editing Condition and Emotion as the main fixed effects and Prompt ID as the random effect. $p > F$ is the significance level of the fixed effect ANOVA. $R^2$ is the model's coefficient of determination to indicate goodness-of-fit.**

| Response | Linear Effects Model | $p > F$ | $R^2$ |
|---|---|---|---|
| Image- Emotion Alignment (CLIP Score) | Prompt Editing Condition<br>Emotion | <.0001<br><.0001 | .766 |